\documentclass[%
prx,letterpaper,
superscriptaddress,
 amsmath,amssymb,
 aps,nofootinbib,twocolumn,floatfix
]{revtex4-2}

\usepackage{graphicx}
\usepackage{dcolumn}
\usepackage{bm}
\usepackage{physics}
\usepackage{sidecap}
\usepackage{float}
\usepackage{scalerel}
\usepackage{xr}
\usepackage{multibib}
\usepackage{xcolor}
\newcites{methods}{References}
\setcitestyle{super}
\makeatletter

\begin{document}
 \title{Quantum thermalization and Floquet engineering in a spin ensemble with a clock transition} 
 

\author{Mi Lei}	
	\affiliation{Kavli Nanoscience Institute and Thomas J. Watson, Sr., Laboratory of Applied Physics, California Institute of Technology, Pasadena, California 91125, USA}
	\affiliation{Institute for Quantum Information and Matter, California Institute of Technology, Pasadena, California 91125, USA}
\author{Rikuto Fukumori}	
	\affiliation{Kavli Nanoscience Institute and Thomas J. Watson, Sr., Laboratory of Applied Physics, California Institute of Technology, Pasadena, California 91125, USA}
	\affiliation{Institute for Quantum Information and Matter, California Institute of Technology, Pasadena, California 91125, USA}
 \author{Chun-Ju Wu}	
	\affiliation{Kavli Nanoscience Institute and Thomas J. Watson, Sr., Laboratory of Applied Physics, California Institute of Technology, Pasadena, California 91125, USA}
	\affiliation{Institute for Quantum Information and Matter, California Institute of Technology, Pasadena, California 91125, USA}

\author{Edwin Barnes}
	\affiliation{Department of Physics, Virginia Technology, Blacksburg, VA 24061, USA}
\author{Sophia Economou}
	\affiliation{Department of Physics, Virginia Technology, Blacksburg, VA 24061, USA}
\author{Joonhee Choi}
        \email[]{joonhee.choi@stanford.edu}
	\affiliation{Department of Electrical Engineering, Stanford University, Stanford, California 94305, USA}
\author{Andrei Faraon}
 	\email[]{faraon@caltech.edu}
	\affiliation{Kavli Nanoscience Institute and Thomas J. Watson, Sr., Laboratory of Applied Physics, California Institute of Technology, Pasadena, California 91125, USA}
	\affiliation{Institute for Quantum Information and Matter, California Institute of Technology, Pasadena, California 91125, USA}

\begin{abstract}
Studying and controlling quantum many-body interactions is fundamentally important for quantum science and related emerging technologies. Optically addressable solid-state spins offer a promising platform for exploring various quantum many-body phenomena due to their scalability to a large Hilbert space. However, it is often challenging to probe many-body dynamics in solid-state spin systems due to large on-site disorder and undesired coupling to the environment. Here, we investigate an optically addressable solid-state spin system comprising a strongly interacting ensemble of millions of ytterbium-171 ions in a crystal. Notably, this platform features a clock transition that gives rise to pure long-range spin-exchange interactions, termed the dipolar XY model. Leveraging this unique feature, we investigate quantum thermalization by varying the relative ratio of interaction strength to disorder, dynamically engineering the XY model into other many-body Hamiltonian models, and realizing a time-crystalline phase of matter through periodic driving. Our findings indicate that an ensemble of rare-earth ions serves as a versatile testbed for many-body physics and offers valuable insights for advancing quantum technologies.
\end{abstract}

\maketitle

Many-body quantum phenomena arise when multiple particles interact, playing a crucial role in contemporary physics and engineering \cite{eisert2015quantum,tasaki2020physics}. Among the systems exhibiting many-body physics, optically addressable solid-state spins have become a particularly rich platform to study due to their scalability to very large numbers of spins \cite{awschalom2018quantum} and their potential applications in quantum simulations \cite{georgescu2014quantum}, sensing \cite{degen2017quantum,rovny2024new}, and information processing \cite{divincenzo2000physical}.

To achieve versatility in quantum applications, it is essential to manipulate the time evolution of a many-body system in a programmable manner \cite{fauseweh2024quantum}. However, achieving local individual control of spins in solid-state systems is often challenging due to their nanometric proximity. To address this issue, a global control sequence--composed of either pulsed or continuous driving--can be applied periodically to the entire system, effectively manipulating many-body dynamics\cite{shirley1965solution, choi2020robust}. This periodic driving method, known as Floquet Hamiltonian engineering, enables the efficient realization of various many-body Hamiltonians with different interaction types and strengths, transformed from the original system Hamiltonian~\cite{goldman2014periodically, bordia2017periodically, bluvstein2021controlling, geier2021floquet}. Notably, periodically driven quantum systems can also exhibit a variety of exotic non-equilibrium phases of matter, such as Discrete Time Crystals (DTCs)\cite{else2016floquet,khemani2016phase,yao2018time,zhang2017observation,choi2017observation}, stabilized by many-body interactions.

In this context, a strongly interacting ensemble of spins with robust global control protocols can be employed to explore many-body physics and out-of-equilibrium dynamics. Rare-earth ions (REIs) doped in solids emerge as promising candidates for a many-body testbed due to their highly coherent optical and spin transitions at cryogenic temperatures, scalability, and ease of integration into photonic devices \cite{thiel2011rare, lei2023many, ourari2023indistinguishable}. Moreover, the diversity in REI species, hosts, and concentrations, combined with naturally arising dipole-dipole interactions~\cite{merkel2021dynamical} or engineered cavity-mediated interactions~\cite{lei2023many}, introduces novel experimental control parameters and offers theoretical insights for exploring large-scale quantum many-body systems and their dynamics. However, despite numerous spectroscopic studies on large ensembles of REIs that have focused on identifying materials with high coherence times \cite{bottger2006optical, thiel2012optical} for quantum technologies such as quantum transducers \cite{williamson2014magneto} and memories \cite{businger2022non}, less effort has been devoted to the microscopic understanding, control, and engineering of these systems.

\begin{figure}
    \centering
    \includegraphics[width=\linewidth]{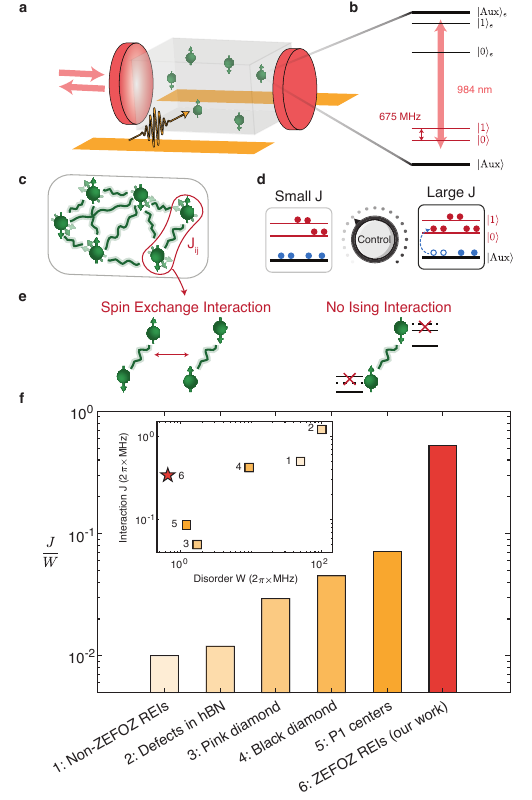}
    \caption{\textbf{Many-body platform based on rare-earth ions.} \textbf{a,} An ensemble of $\approx$10$^6$ rare-earth ion spins (green arrows) is coupled to an optical cavity (red mirrors) and a coplanar waveguide (gold stripes). The optical input and output (red arrows) are sent from the same port for spin initialization and readout. \textbf{b,} The energy levels of each spin, with four ground states ($\ket{0}$, $\ket{1}$, and $\ket{\text{Aux}}$) and four optically excited states ($\ket{0}_e$, $\ket{1}_e$, and $\ket{\text{Aux}}_e$). Both $\ket{\text{Aux}}$ and $\ket{\text{Aux}}_e$ are doubly degenerate states, whereas $\ket{0}$, $\ket{0}_e$, $\ket{1}$, and $\ket{1}_e$ are clock states. We define $\ket{0}$ and $\ket{1}$ as a qubit and read out their state via the optical transition around $984$~nm. \textbf{c,} An interacting spin ensemble where spins $i$ and $j$ interact via pairwise interactions with strength $J_{ij}$. 
    \textbf{d,} Control of the average interaction strength $J$ by varying the population distribution between the qubit manifold $\{\ket{0}, \ket{1}\}$, and the auxiliary states $\ket{\text{Aux}}$ via optical pumping (Supplementary Information). \textbf{e,} Our spin system permits pure spin-exchange interactions with no Ising interactions. \textbf{f,} Benchmarking our many-body platform to other solid-state electronic spin systems. The bar chart shows the ratio of the average interaction strength to disorder, $J/W$, for different systems: 1: REIs with nonzero first-order-Zeeman shift (Non-ZEFOZ REIs) \cite{gupta2023robust, xie2021characterization}; 2: Defects in hexagonal boron nitride (hBN) \cite{gong2023coherent}; 3: Pink diamond \cite{he2023quasi,he2024experimental}; 4: Black diamond \cite{choi2017depolarization}; 5: P1 centers \cite{davis2023}; 6: ZEFOZ REIs (our work). Inset: comparison of the absolute values of $J$ and $W$.} 
    \label{fig1}
\end{figure}

In this work, we report on the characterization and control of quantum many-body dynamics in a dense ensemble of approximately $10^6$ ytterbium-171 ions in a nanophotonic cavity with a yttrium orthovanadate host crystal ($^{171}$Yb$^{3+}$:YVO$_4$). These high-density REIs are randomly positioned with an average distance of $\approx$$9$~nm (or equivalently, $\approx$$86$~ppm in concentration) in an effective three-dimensional volume defined by the cavity mode (Fig.~\ref{fig1}a). Each Yb ion provides both microwave and optical transitions for spin state manipulation and state-selective readout, respectively. The nanophotonic optical cavity enables high-fidelity initialization and fast readout of Yb ions, while coherent microwave control is achieved via a coplanar waveguide. Specifically, the ground and optically excited states of Yb comprise $\{\ket{0}, \ket{1}, \ket{\text{Aux}} \}$ and $\{\ket{0}_e, \ket{1}_e, \ket{\text{Aux}}_e\}$, respectively, resulting from the hybridization of its electron and nuclear spins (Fig.~\ref{fig1}b) \cite{kindem2018}. In this study, the spin transitions are defined within the ground state manifold using $\{\ket{0}, \ket{1}\}$, serving as an effective ``qubit'' with a microwave transition frequency of 675~MHz. For spin state readout, we utilize the cavity-enhanced, resonant optical transition between $\ket{1}$ and $\ket{0}_e$ at a wavelength of 984 nm, enabling the optical detection of state-selective photoluminescence signals (Supplementary Information).

In essence, the goal of studying many-body physics in an experiment is to observe coherent quantum phenomena over extended durations governed by the target unitary dynamics of the system, while minimizing undesired incoherent coupling to external environments. In this regard, our REI platform offers distinct advantages over other solid-state spin systems for exploring the dynamics of a {\it closed} many-body quantum system, as outlined below.

First, the chosen spin states, $\{\ket{0}, \ket{1}\}$, are first-order insensitive to external electromagnetic fluctuations at zero magnetic fields, known as the ``clock'' transition~\cite{kindem2018}. Consequently, the spins are less susceptible to both decoherence and inhomogeneity induced by the external environment. Second, the electronic spin $g$ factor, which determines the strength of the dipole moment, is approximately three times higher than that of a single electron~\cite{ruskuc2022nuclear}. This results in stronger dipole-dipole interactions between REIs (Fig.~\ref{fig1}c). Third, we can control the overall interaction strength of the system, $J$, defined using the average nearest-neighbor distance (Supplementary Information), by adjusting the effective density of REIs within the qubit manifold $\{\ket{0}, \ket{1}\}$ (Fig.~\ref{fig1}d). This manipulation is achieved by varying the population distribution between the qubit manifold $\{\ket{0}, \ket{1}\}$ and the nonparticipating auxiliary ground states $\ket{\text{Aux}}$ through optical initialization (Supplementary Information). Lastly, unlike conventional dipole-dipole interactions, which include both spin-exchange and energy-shifting Ising interactions, our spin system realizes a pure spin-exchange interaction without the Ising component (Fig.~\ref{fig1}e). This Hamiltonian is known as the dipolar XY model \cite{hazzard2014quantum}, which has recently been experimentally investigated in various platforms to explore fundamental many-body phenomena and their applications, such as continuous symmetry breaking \cite{chen2023continuous} and spin squeezing \cite{bornet2023scalable}. The naturally occurring XY model in three-dimensional dipolar systems offers a novel and complementary configuration for studying similar physics in the solid-state setting. 

The combination of strong spin-spin interactions, weak coupling to the external bath, cavity-enhanced fast readout, and effective spin density control positions our system as a promising platform for studying many-body spin dynamics. Notably, our spin system features the highest ratio of interaction strength to transition inhomogeneity among solid-state electronic spin systems, underscoring its significance in probing many-body dynamics where collective interaction effects are minimally hindered by on-site disorder (Fig.~\ref{fig1}f). 

\begin{figure*}
    \centering
    \includegraphics[width=\linewidth]{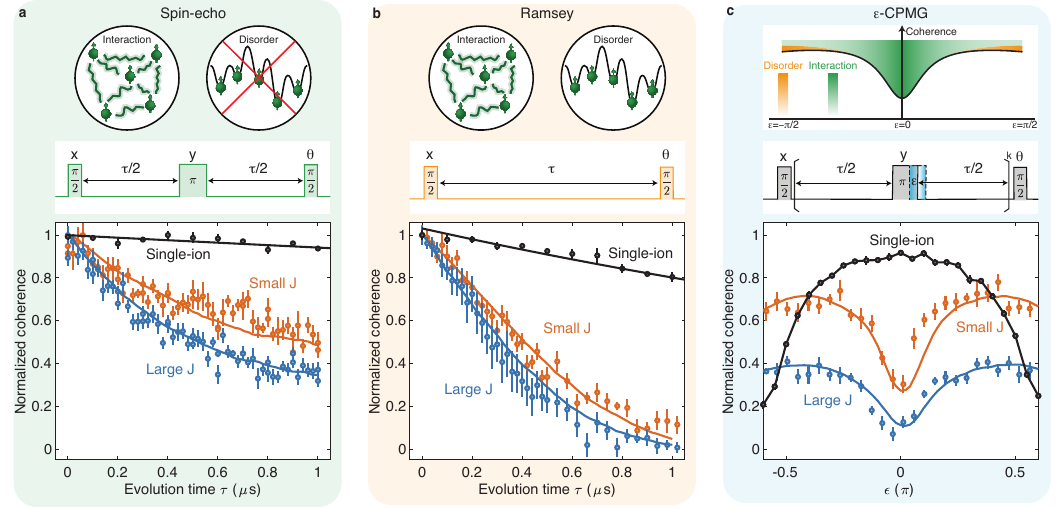}
    \caption{\textbf{Characterization of decoherence dynamics.} We characterize spin system decoherence arising from interaction and disorder using spin-echo, Ramsey, and $\epsilon$-CPMG sequences. We compare three different cases: an interacting spin ensemble with large $J$ (blue) and small $J$ (orange), and an isolated single ion as a reference (black). \textbf{a,} In the spin-echo measurement, spin-spin interactions drive decoherence with decay rates dependent of $J$, while disorder is decoupled to the zeroth order. \textbf{b,} In the Ramsey measurement, both interaction and disorder contribute to decoherence. In \textbf{a} and \textbf{b}, coherence is measured as a function of the free evolution time, $\tau$, and normalized to their respective maximum coherence. The last $\pi/2$ pulse with a variable rotation axis $\theta$ is used to extract coherence from photoluminescence signals (Supplementary Information). \textbf{c}, In the $\epsilon$-CPMG measurement, a $\pi+\epsilon$ pulse is applied $k$ times at a fixed $\tau$. Coherence is measured as a function of the rotation angle offset, $\epsilon$, and then normalized to the maximum coherence observed in the spin-echo measurement. Note that $\epsilon$ can control the decoherence contributions caused by disorder and interaction. The interacting spin ensemble reveals unconventional behavior where coherence is maximized when $\epsilon \approx \pm \pi/2$. The error bars are obtained from fits to the experimental data, and simulations are shown as solid lines (Supplementary Information). $\tau = 300$ ns and $k = 8$ are chosen for both the large and small $J$ cases, while $\tau = 5.8$ $\mu$s and $k = 400$ are chosen for the single ion. In \textbf{c}, we rescale the simulation results by an $\epsilon$-independent prefactor $<1$ to match the experimental data.}
    \label{fig2}
\end{figure*}
\section*{Characterization of a strongly interacting spin system}
The very first step before we utilize our REI spin system as a many-body testbed is to quantitatively characterize the interaction and disorder strengths in the system. To this end, we describe the dynamics of our interacting spin system using the following Hamiltonian (in a rotating frame), $\hat{H}$, defined within the qubit manifold $\{\ket{0},\ket{1}\}$:
\begin{equation} \label{H}
    \hat{H} = \hat{H}_\text{dis}+\hat{H}_\text{int}
\end{equation}
where $\hat{H}_\text{dis} = \sum_i^N\Delta_i \hat{S}_z^i$ is the on-site disorder Hamiltonian with spin detuning $\Delta_i$ for ion $i$, and $\hat{H}_\text{int} = \sum_{ij, \; i>j}^N J_{ij}(\hat{S}_x^i \hat{S}_x^j+ \hat{S}_y^i \hat{S}_y^j) = \sum_{ij, \; i>j}^N \frac{J_{ij}}{2}(\hat{S}_+^i \hat{S}_-^j+ \hat{S}_-^i \hat{S}_+^j)$ is the long-range, dipolar spin-exchange Hamiltonian with position- and orientation-dependent pairwise interaction strength $J_{ij}$ between two ions $i$ and $j$ (see Supplementary Information for details). Here, $\hat{S}_\mu^i$ is the spin-1/2 operator of ion $i$ along the $\mu$-axis ($\mu=x,y,z$), and $\hat{S}_{\pm}^i = \hat{S}_x^i \pm i\hat{S}_y^i$ are the creation and annihilation operators for spin excitation of ion $i$. We define the strength of on-site disorder, $W$, as the full width at half maximum of the probability distribution function of $\Delta_i$.

In strongly interacting spin systems, the decoherence of a spin ensemble is influenced not only by the random, inhomogeneous on-site fields but also by interactions with the rest of the spin system, which act as an intrinsic bath~\cite{kucsko2018critical}. To isolate the decoherence effect arising from spin-spin interactions, we employ the celebrated spin-echo sequence~\cite{hahn1950spin}, which effectively cancels out the on-site disorder in the Hamiltonian at short times (Fig.~\ref{fig2}a). Here, we observe rapid decay of ensemble coherence within $\approx$$1~\mu s$, which is much shorter than that of an isolated single spin independently measured from a reference sample. We confirm that the decay time is highly dependent on the effective spin density, where the decoherence rate increases as $J$ increases (orange/blue markers, Fig.~\ref{fig2}a; see Supplementary Information for additional experimental data). This implies that intrinsic spin-spin interactions dominate the decoherence mechanisms.

To extract the effective spin density and the corresponding average interaction strength, $J$, we conduct numerical simulations based on a closed many-body system (Supplementary Information). We find that the simulations show good agreement with the experiment, revealing spin concentrations of $\approx$46 ppm and $\approx$25 ppm for the cases with large $J \approx 2\pi \times 0.35$ MHz and small $J \approx 2\pi \times 0.19$ MHz, respectively (orange/blue lines, Fig.~\ref{fig2}a). These spin densities within the qubit manifold are reasonable given the total spin densities of all ground states of $\approx$86 ppm, as independently measured by mass spectrometry~\cite{lei2023many}.

Having characterized the interaction strengths, we now focus on identifying the on-site disorder strength, $W$, using the disorder-sensitive Ramsey sequence~(Fig.~\ref{fig2}b). Experimental data show that the Ramsey signal decays faster than the spin-echo signal in both the small and large $J$ regimes, due to the additional contribution of disorder-induced decoherence. By comparing the experimental data to the corresponding Ramsey sequence simulation, we estimate an on-site disorder strength of $W \approx 2\pi \times 0.65$ MHz, which is independent of interaction strengths (orange/blue lines, Fig.~\ref{fig2}b).

To further corroborate the significance of spin-spin interactions in our system, we employ the so-called $\epsilon$-CPMG sequence~\cite{schenken2023long}, a variant of the conventional CPMG sequence (Fig.~\ref{fig2}c). Specifically, the $\epsilon$-CPMG sequence purposely uses an \textit{imperfect} echo pulse with a rotation angle of $\pi + \epsilon$ with nonzero $\epsilon$. For either an isolated spin or a non-interacting spin ensemble affected only by disorder, such imperfect spin rotation with a sizable $\epsilon$ results in non-ideal dynamical decoupling, leading to rapid decoherence compared to $\epsilon = 0$ (black markers, Fig.~\ref{fig2}c). In contrast, we observe drastically different behavior in our high-density spin system, where nonzero $\epsilon$ pulses better preserve ensemble coherence (orange/blue markers, Fig.~\ref{fig2}c).

This feature originates from the different sensitivities of the $\epsilon$-CPMG sequence to on-site disorder and many-body interactions (Supplementary Information). Specifically, when $\epsilon = 0$, spin-spin interactions still lead to rapid decoherence of the spin ensemble because the $\pi$ pulses do not alter their spin-exchange interaction Hamiltonian,  $\hat{H}_\text{int}$ (whereas the effects from the on-site disorder Hamiltonian, $\hat{H}_\text{dis}$, can be decoupled). However, when $\epsilon = \pm \pi/2$, effectively corresponding to a $\pi/2$ pulse, $\hat{H}_\text{int}$ transforms into a Hamiltonian that is the sum of the Ising interaction along the $y$-axis and the Heisenberg interaction (Supplementary Information). The initially prepared spins along the $y$-axis then become an eigenstate under this Hamiltonian. Consequently, using $\pi/2$ pulses instead of $\pi$ pulses in the $\epsilon$-CPMG sequence significantly extends the coherence of the interacting spin ensemble (Supplementary Information). Using the Hamiltonian parameters extracted from the Ramsey and echo sequences, we reproduce the coherence dependence on $\epsilon$ that matches experimental data (orange/blue lines, Fig.~\ref{fig2}c). 

\section*{Microscopic understanding of decoherence mechanisms} 

\begin{figure}[b!]
    \centering
    \includegraphics[width=\linewidth]{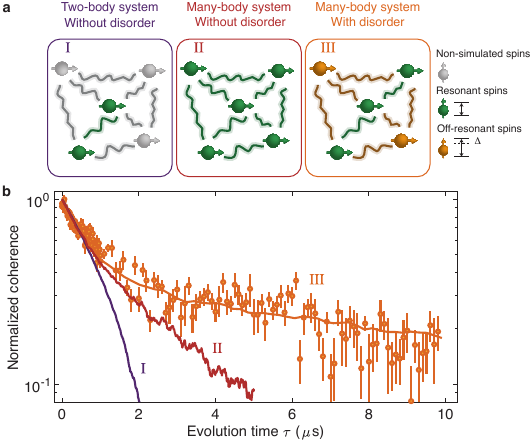}
    \caption{\textbf{Microscopic understanding of decoherence mechanisms.} \textbf{a,} Schematic of Models I, II, and III for spin-echo sequence simulations. Model I considers only the pair of spins with the largest interaction strength and excludes disorder. Models II and III involve an ensemble of $N$ spins with many-body interactions; Model II excludes disorder, while Model III includes disorder with strength $W = 2\pi \times 0.65$~MHz (calibrated from Fig.~2b). \textbf{b,} Comparison of the experimental spin-echo data (markers) against the three models. The case with interaction strength $J \approx 2\pi \times 0.19$~MHz is considered for comparison. In each realization of the Monte Carlo numerical simulations, $N=9$ spins are randomly positioned based on the lattice structure and the given spin density. The interaction strengths between each pair of spins are calculated based on their positions (Supplementary Information). The error bars are obtained from fits to the experimental data, and simulations are shown as solid lines.}
    \label{fig3}
\end{figure}

When probing the spin-echo dynamics over longer timescales, we observe that the decoherence profile starts deviating from a simple exponential decay, displaying much slower relaxation at later times (orange markers, Fig.~\ref{fig3}b). Remarkably, a numerical simulation with calibrated disorder and interaction strengths (Model III) shows excellent agreement with the experimental data even at longer timescales (orange line, Fig.~\ref{fig3}b). To comprehend these late-time observations, we further consider two simpler theoretical models, I and II, which consider two-body and many-body systems with {\it no} disorder, respectively (Fig.~\ref{fig3}a; see Supplementary Information for details). We find that both fail to capture the late-time slowdown in decoherence (purple/red lines, Fig.~\ref{fig3}b).

\begin{figure*}
    \centering
    \includegraphics[width=\linewidth]{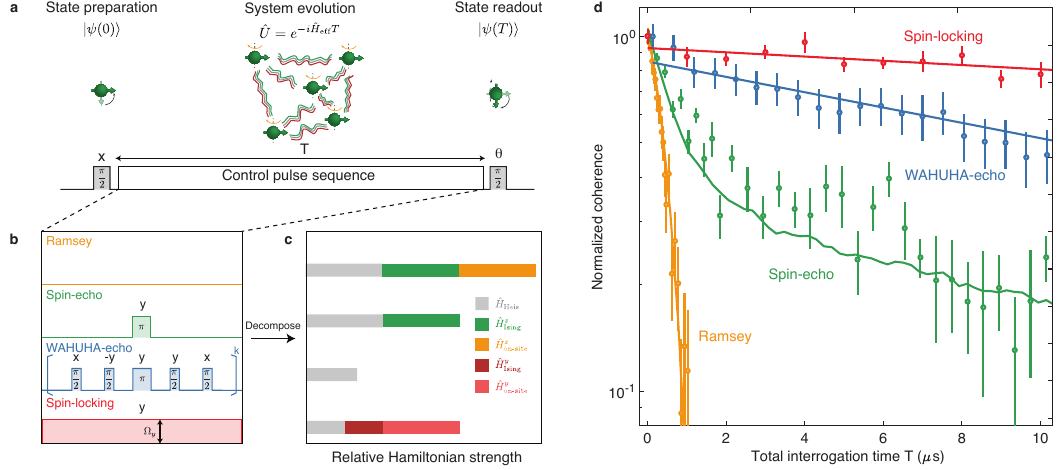}
    \caption{\textbf{Controlling system evolution via dynamic Hamiltonian engineering.} \textbf{a,} Hamiltonian engineering protocol based on control pulse sequences. The first $\pi/2$ pulse around the $x$-axis initializes the system into a globally polarized state along the $y$-axis. The spin states evolve as $\ket{\Psi(T)} = \hat{U} \ket{\Psi(0)}$ over the total interrogation time $T$, where $\hat{U} = e^{-i\hat{H}_\text{eff} T}$ is the time-evolution operator governed by the sequence-dependent effective Hamiltonian, $\hat{H}_\text{eff}$, obtained through averaged Hamiltonian theory (Supplementary Information). The spin polarization along the $y$-axis is extracted by fitting the photoluminescence signal as a function of the rotation axis $\theta$ of the last $\pi/2$ pulse (Supplementary Information). \textbf{b,} Ramsey, spin-echo, WAHUHA-echo, and spin-locking sequences for dynamic Hamiltonian engineering. A Rabi frequency of $\Omega_y \approx 2\pi \times 10~$MHz is used in the spin-locking sequence. The base sequence period of the WAHUHA-echo sequence is 132~ns, limited by finite pulse durations (Supplementary Information). \textbf{c,} We decompose the effective Hamiltonian of each sequence into a sum of different Hamiltonians (see Eq.~(\ref{Heffmain})). 
    The relative weights between the different Hamiltonian terms are represented by the lengths of the bars in the plot. See the main text for details. \textbf{d,} Comparison of decoherence profiles as a function of $T$ under different control sequences. Coherence is normalized to the maximum coherence for each sequence, with error bars indicating the standard deviation of the experimental data. Solid lines represent simulations for the spin-echo and Ramsey sequences, while single-exponential fits are applied to the spin-locking and WAHUHA-echo sequences.}
    \label{fig4}
\end{figure*}

Notably, the cross-over from fast to slow decoherence occurs after a time approximately given by the inverse disorder strength, $\tau \sim 1/W$ (Fig.~\ref{fig3}b). We attribute this to the breakdown of perfect decoupling of on-site disorder when the free evolution time becomes comparable to the inverse disorder strength, implying the critical role of on-site disorder in late-time many-body dynamics. Specifically, high-order disorder effects, arising from the non-zero commutator between the disorder and interaction Hamiltonians ($[\hat{H}_\text{dis},\hat{H}_\text{int}] \neq 0$), become pronounced at late times, imposing energetic penalties on spin-exchange processes and thereby slowing down the decoherence rate (Supplementary Information).

Our microscopic analysis allows us to dissect the underlying decoherence mechanisms of the interacting spin system influenced by both interactions and disorder. Moreover, the analysis strongly supports an exceptionally high degree of coherence in a many-body regime, as evidenced by its excellent agreement with simulations based on a closed many-body system.

\section*{Control of system evolution via dynamic Hamiltonian engineering} 
Having established that the decoherence dynamics of an individual spin are governed by interactions within the system, we use the decoherence profile as a proxy to explore dynamically engineered many-body Hamiltonians with control pulse sequences (Fig.~\ref{fig4}a-c). Intuitively, each spin can be viewed as a quantum sensor sensitive to interactions with surrounding spins, experiencing interaction-induced dephasing, which is often termed quantum thermalization~\cite{martin2023controlling}. 

Specifically, we prepare an initial state in a globally polarized state along the $y$-axis, let the spin system evolve under a control pulse sequence over an interrogation time $T$, and apply a $\pi/2$ pulse with a variable rotation axis $\theta$ at the end of the sequence to measure the mean coherence of all individual spins. Through this experiment, we can probe the decoherence dynamics of a spin ensemble as a function of $T$, defined as the average spin polarization along the $y$-axis, $P(T) = \frac{1}{N} \bra{\Psi(T)}\hat{S}^\text{tot}_y\ket{\Psi(T)}$. Here, $\hat{S}^\text{tot}_y=\sum_{i}^N\hat{S}_y^i$ is  the total spin operator along the $y$-axis with $N$ being the total number of spins, and  $\ket{\Psi(T)} = e^{-i \hat{H}_\text{eff} T} \ket{\Psi(0)}$ is the quantum state at time $T$ obtained through time evolution under the dynamically-engineered effective Hamiltonian $\hat{H}_\text{eff}$.

Crucially, with dynamic Hamiltonian engineering of the original dipolar XY Hamiltonian, we can realize a wide class of different effective Hamiltonians, $\hat{H}_\text{eff}$, parameterized as
\begin{align} \label{Heffmain}
    \hat{H}_\text{eff} &= w_\text{Heis} \hat{H}_\text{Heis} + \sum_{\mu=x,y,z} \left(w_\text{on-site}^\mu \hat{H}_\text{on-site}^\mu + w_\text{Ising}^\mu \hat{H}_\text{Ising}^\mu \right).
\end{align}
Here, $w_\text{Heis}$, $w_\text{on-site}$, and $w_\text{Ising}$ are relative weights between the different Hamiltonians: the Heisenberg Hamiltonian $\hat{H}_\text{Heis}=\sum_{ij, \; i>j}^N J_{ij}\vec{S}^i$$\boldsymbol{\cdot}$$\vec{S}^j$, the on-site Hamiltonian $\hat{H}_\text{on-site}^\mu = \sum_i^N h_\mu^i \hat{S}_\mu^i$, and the Ising Hamiltonian $\hat{H}_\text{Ising}^\mu = \sum_{ij, \; i>j}^N J_{ij}\hat{S}_\mu^i \hat{S}_\mu^j$, where $\vec{S}^i = (\hat{S}_x^i, \hat{S}_y^i, \hat{S}_z^i)$ is a vectorized spin operator, and $\vec{h}^i = (h_x^i, h_y^i, h_z^i)$ is an effective on-site field for ion $i$. Note that the Heisenberg Hamiltonian $\hat{H}_\text{Heis}$ is isotropic, whereas the other two Hamiltonians exhibit directionality along the $\mu$-axis. 

As depicted in Figs.~\ref{fig4}b and \ref{fig4}c, we implement and compare four different effective Hamiltonians using Ramsey, spin-echo, Waugh-Huber-Haberlen (WAHUHA)-echo \cite{waugh1968approach}, and spin-locking sequences. The on-site Hamiltonian is determined by the inhomogeneous disordered field along the $z$-axis ($\vec{h}^i = (0, 0, \Delta_i)$) for the Ramsey and spin-echo sequences, and the homogeneous control field along the $y$-axis with strength $\Omega_y$ ($\vec{h}^i = (0, \Omega_y, 0)$) for the spin-locking sequence. Meanwhile, the WAHUHA-echo sequence is designed to realize the pure Heisenberg Hamiltonian (to leading order), thereby protecting initially polarized spins from interaction-induced decoherence. This protection arises because a globally polarized spin state is an eigenstate of the Heisenberg Hamiltonian \cite{choi2020robust}. Additionally, the spin-locking sequence prevents spin dephasing by employing a strong pinning field along the $y$-axis with strength $\Omega_y \gg J$ after the initial state preparation, effectively arresting spin dynamics.

In line with these theoretical expectations, both the WAHUHA-echo and spin-locking sequences demonstrate significantly prolonged coherence times compared to the Ramsey and spin-echo sequences, confirming the effective engineering of the underlying many-body Hamiltonian (Fig.~\ref{fig4}d). However, we observe that the coherence of the WAHUHA-echo and spin-locking sequences still decays over time, with $1/e$ time constants of $\approx$$20$ $\mu s$ and $\approx$$73$ $\mu s$, respectively. We attribute these decays to imperfections in spin ensemble control caused by finite pulse duration and rotation angle errors, as well as fast time-dependent fluctuations of on-site fields originating from the external spin bath (Supplementary Information). These issues could be addressed by implementing a more robust control sequence and improving control pulse fidelity \cite{choi2020robust}.

\section*{Signatures of discrete time-crystalline phase} 
\begin{figure}
    \centering
    \includegraphics[width=\linewidth]{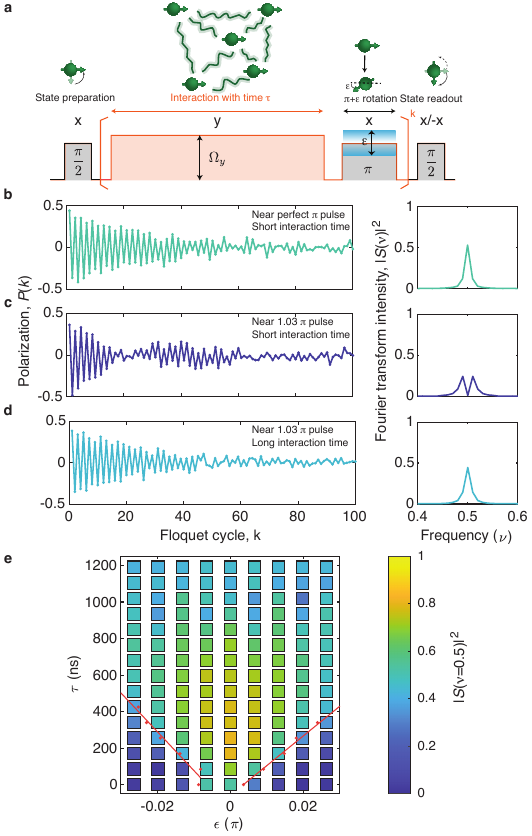}
    \caption{\textbf{Experimental signatures of discrete time-crystalline states.} \textbf{a,} Floquet pulse sequence for probing DTC states. All spins are initially polarized along the $y$-axis by the first $\pi/2$ pulse around the $x$-axis, followed by the periodic repetition of a base sequence with periodicity $\tau$, applied $k$ times. The base sequence contains spin-locking along the $y$-axis for a duration $\tau$ and a global spin rotation by an angle of $\pi+\epsilon$ around the $x$-axis (ignoring the finite spin-rotation duration). The photoluminescence signals, $C_+$ and $C_-$, are collected under the last $\pi/2$ pulse with opposite phases, $x$ and $-x$, respectively. We estimate the normalized spin polarization, $P(k) = \frac{|C_{+}-C_{-}|}{|C_{+}+C_{-}|}$, at stroboscopic times $t=k\tau$. \textbf{b-d,} The polarization, $P(k)$, as a function of Floquet cycle $k$, and the corresponding Fourier spectrum, $|S(\nu)|^2$, are shown for various values of $\tau$ and $\epsilon$: \textbf{b}, $\tau \approx 0$, $\epsilon\approx0$; \textbf{c}, $\tau \approx 0$, $\epsilon\approx0.03\,\pi$; \textbf{d}, $\tau \approx 425$~ns, $\epsilon\approx0.03\,\pi$. The subharmonic oscillation is observed in \textbf{d}, despite the nonzero angle offset $\epsilon$, attributed to the stabilization of DTC states by spin-spin interactions. \textbf{e,} DTC phase diagram constructed using the subharmonic peak intensity at $\nu=0.5$, i.e., $|S(\nu=0.5)|^2$. We observe characteristic linear phase boundaries (dashed lines), determined by identifying the critical perturbation strength (markers) where $|S(\nu=0.5)|^2 < 0.4$ (Supplementary Information).}
    \label{fig5}
\end{figure}
Studying non-equilibrium dynamics under periodic driving unveils the microscopic mechanisms of driven many-body phenomena, such as DTCs~\cite{else2016floquet,khemani2016phase,yao2018time,zhang2017observation,choi2017observation}, revealing an interplay among disorder, dimensionality, interactions, and robustness to both intrinsic and extrinsic perturbations. The canonical model for observing DTCs relies on many-body localization (MBL) and disordered Ising interactions, which lead to the emergence of symmetry-broken spin-glass ordering~\cite{mi2022time}. Since our system Hamiltonian consists of pure spin-exchange interactions, we employ a spin-locking sequence to effectively realize an Ising interaction along the $y$-axis (Fig.~\ref{fig5}a). In particular, the spin-locking sequence enables long-lived many-body coherence, providing an ideal setting to study the emergent dynamics at late times under periodic driving (Supplementary Information). While we do not expect our three-dimensional spin system to exhibit MBL, we investigate the possibility of inducing time-crystalline behavior in a disordered system, akin to the critical DTCs where the relaxation of spin ordering occurs exponentially slowly \cite{ho2017critical}.

Specifically, we apply a Floquet pulse sequence for probing DTCs as follows: we first prepare the globally polarized spin state along the $y$-axis, immediately drive the polarized spins along the same direction for a variable spin-locking duration $\tau$, rotate every spin by an angle $\pi + \epsilon$ around the $x$-axis, and repeat this ``spin-lock-and-rotate'' operation $k$ times (Fig.~\ref{fig5}a). Subsequently, using the last $\pi/2$ pulse with two opposite phases of $x$ and $-x$, we measure the normalized spin polarization, ${P}(k)$, at stroboscopic times $t=k\tau$, where $k$ is an integer. The repeated base sequence imposes a discrete time-translation symmetry with a period of $\tau$, during which spins are allowed to interact (ignoring the finite spin-rotation duration), providing the desired setting to test whether our many-body system can break this discrete time-translation symmetry.

Experimentally, we find that when $\epsilon \approx 0$, corresponding to near-perfect $\pi$ rotations, with a short interaction period $\tau \approx 0$, the spin polarization, $P(k)$, oscillates up and down over time, exhibiting a $2\tau$-periodic oscillation (Fig.~\ref{fig5}b). The decay envelope of the oscillation is caused by imperfections in the $\pi$ pulse, as well as the dephasing of the spin-locking signal itself. The discrete Fourier transform (DFT) of the oscillating $P(k)$ reveals a subharmonic peak at a frequency of $\nu = \frac{1}{2}$ in units of the base period $\tau$. However, this subharmonic peak arises trivially from the finetuned condition of $\epsilon = 0$; as soon as we introduce a non-zero systematic rotation angle error, $\epsilon \neq 0$, as a perturbation to the system, the subharmonic response is disrupted, and instead, the system exhibits a beat note where the corresponding DFT spectrum shows $\epsilon$-dependent frequency splitting (Fig.~\ref{fig5}c). Crucially, however, the subharmonic oscillation of the polarization can be restored by allowing for a longer interaction time of $\tau \approx 0.145 \times \frac{2\pi}{J} \approx 425 $~ns at a nonzero $\epsilon$, which is indicative of a DTC phase exhibiting robustness against perturbations (Fig.~\ref{fig5}d). 

We proceed to investigate the stability of this subharmonic behavior more systematically for various values of interaction time $\tau$ and perturbation strength $\epsilon$ by constructing a phase diagram using the subharmonic peak intensity at $\nu=0.5$ (Fig.~\ref{fig5}e). The resulting DTC phase diagram shows a characteristic linear phase boundary (red dashed lines, Fig.~\ref{fig5}e), consistent with observations from other many-body platforms \cite{zhang2017observation,choi2017observation}. We further substantiate the robustness of the observed DTC phases by confirming the persistent subharmonic oscillations when varying the initial spin states through global rotations away from the $y$-axis (Supplementary Information).

\section*{Conclusion and outlook} 
Our experimental demonstrations showcase that a REI system provides a versatile and flexible testbed for many-body physics. We envision that REI platforms hold greater potential to serve as large-scale analog quantum simulators in the solid state, offering unique features compared to other solid-state spin systems. First, REI platforms offer flexible engineering options, including a variety of fabrication-friendly choices for ion species, host crystals, and doping concentrations spanning a wide dynamic range from ppb to a few percent. Second, co-doping different REI species allows for the simultaneous engineering of two distinct groups of many-body systems, facilitating the study of heterogeneous spin-spin interactions \cite{lephoto2012synthesis, el2017influence, baker2019rare}, analogous to the dual-species experiments using neutral atom arrays ~\cite{singh2022dual}. Third, proximal nuclear spins within the crystal can serve as an additional quantum register. The nuclear spin-spin interactions can be mediated through engineered hyperfine interactions with a Yb ion, offering additional functionality such as a quantum memory \cite{ruskuc2022nuclear}. Lastly, all of these control knobs, combined with on-chip integration using nanotechnology \cite{lei2023many,ourari2023indistinguishable}, enable a scalable REI system, opening up a range of practical applications from quantum simulation to networking and sensing.
\section*{Acknowledgments}
We thank S. L. N. Hermans and T. Xie for reading the paper and providing useful feedback, and A. Ruskuc and A. Beckert for useful discussions. \textbf{Funding:} This work was supported by U.S. Department of Energy, Office of Science, National Quantum Information Science Research Centers, Co-design Center for Quantum Advantage (contract number DE-SC0012704), Institute of Quantum Information and Matter, an NSF Physics Frontiers Center (PHY-1733907) with support from the Moore foundation EPI program, NSF UCLA, NSF QUIC-TAQS (Award Number: 2137984). The initial device and experimental setup development was supported by the Office of Naval Research award no. N00014-19-1-2182 and N00014-22-1-2422. The device nanofabrication was performed in the Kavli Nanoscience Institute at the California Institute of Technology. R.F. acknowledges the support from the Quad fellowship. E.B. acknowledges support from the National Science Foundation (grant no. 1847078). J.C. acknowledges support from the Terman Faculty Fellowship at Stanford. \textbf{Author Contributions:} M.L., J.C., and A.F. conceived the idea and experiment. M.L. built the experimental setup, performed the measurements, and analyzed the data. R.F. performed the numerical simulations. C.J.W. took the data for a reference sample. M.L., R.F., E.B., S.E.E., J.C., and A.F. interpreted the results. M.L., J.C., and A.F. wrote the manuscript with inputs from all authors. All work was supervised by J.C. and A.F. \textbf{Competing interests:} The authors declare no competing interests. \textbf{Data and materials availability:} All data are available in the manuscript or the supplementary information.

\color{black} 
\bibliography{main}

\clearpage

\end{document}


\begin{center}
    {\Large \textbf{Supplementary Information}\par}
\end{center}
\vspace{0.5cm}


\baselineskip18pt


\renewcommand{\thefigure}{S\arabic{figure}}
\renewcommand{\thetable}{S\arabic{table}}
\renewcommand{\theequation}{S\arabic{equation}}

\tableofcontents


\section{Materials and Methods}

\subsection{Experimental setup and device}
\begin{figure}[h!]
    \centering
    \includegraphics[width=\linewidth]{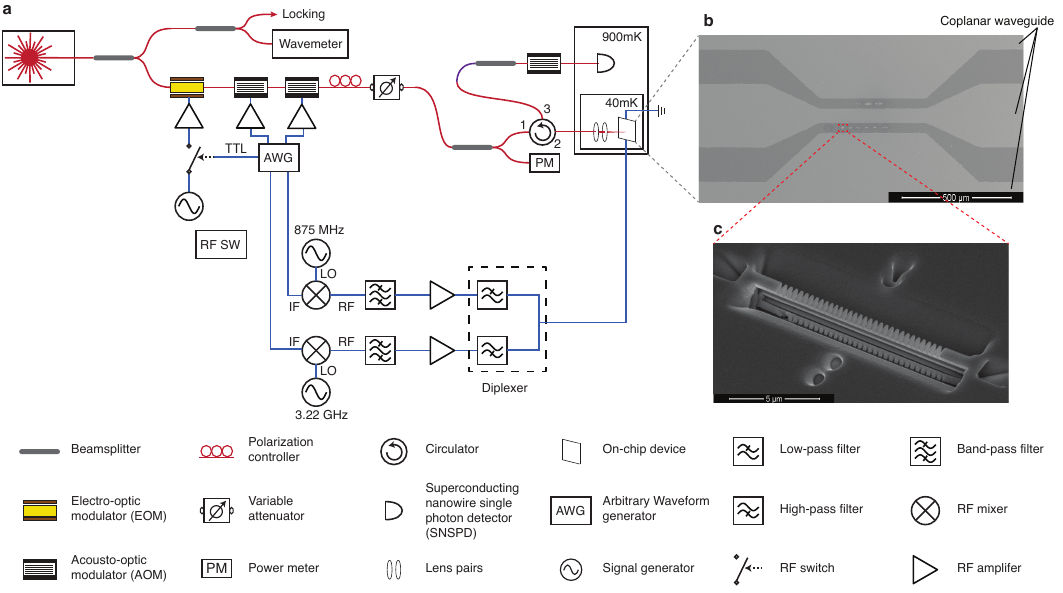}
    \caption{\textbf{Experimental setup and device.} \textbf{a,} The red line indicates the beam path of a 984 nm laser. The laser is locked to the frequency of the A transition. A frequency sideband at the F transition is generated by an EOM. Optical pulses are generated using AOMs. Both the EOM and AOMs are driven by gated RF sources. The light passes through a circulator to the device, and the reflected light is directed to a SNSPD for time-resolved photon counting. The blue lines indicate microwave signal delivery. RF pulses for the ground state spin transition ($f_g \approx 0.675$ GHz) and excited state spin transition ($f_e \approx 3.37$ GHz) are generated using frequency up-conversion from the local oscillator signal mixed with the output of the AWG. The signals after the mixers pass through band-pass filters and amplifiers, and are then combined using a diplexer before being sent to the device. \textbf{b,} Scanning electron microscope image of the chip in the dilution fridge, surrounded by a coplanar waveguide. \textbf{c,} Scanning electron microscope image of the nanophotonic device.}
    \label{setup}
\end{figure}

\begin{figure}
    \includegraphics[width=\linewidth]{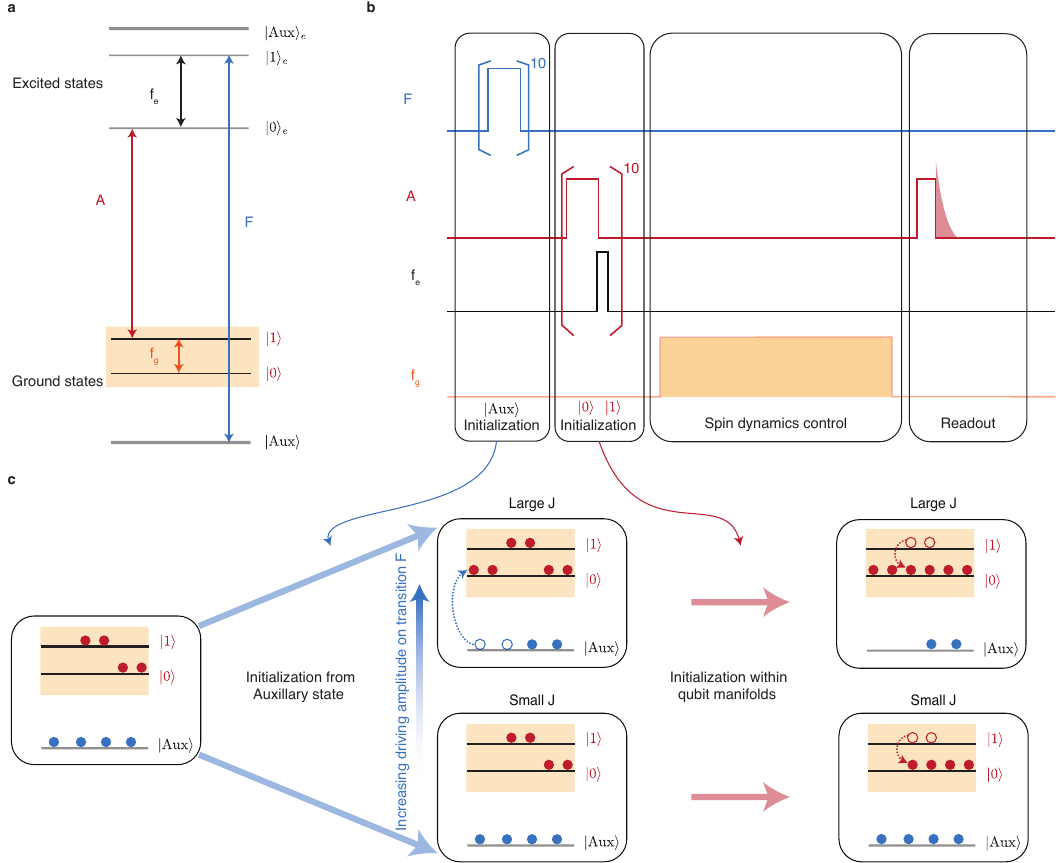}
    \caption{\textbf{Experimental sequences.} \textbf{a,} The simplified energy level diagram for a $^{171}$Yb$^{3+}$ ion in a YVO$_4$ crystal. Both the ground and optically excited states exhibit fine structures labeled as \{$\ket{0}, \ket{1}, \ket{\text{Aux}}$\} and \{$\ket{0}_e, \ket{1}_e, \ket{\text{Aux}}_e$\}, respectively. Microwave spin transitions occur at frequencies $f_g \approx 0.675$ GHz and $f_e \approx 3.37$ GHz for the ground and excited states, respectively. The optical transitions A and F occur near a wavelength of $984$ nm, with a separation of $6.118$ GHz. Note that transition $A$ is coupled to the cavity mode used for fast spin state readout, while transition $F$ is not coupled to the cavity mode and is driven for optical initialization. \textbf{b,} Experimental sequences include state initialization, spin dynamics control, and optical readout. \textbf{c,} Control of the average spin-spin interaction strength $J$ within the qubit manifold \{$\ket{0}, \ket{1}$\} (orange shaded area). Adjusting the driving amplitude on the F transition during initialization effectively controls the transfer of population from the auxiliary state $\ket{\text{Aux}}$ to the qubit manifold \{$\ket{0}, \ket{1}$\} (blue arrows). Subsequently, a combination of optical pulses driving the $A$ transition and microwave pulses driving the $f_g$ transition is applied to polarize the spin state to $\ket{0}$ (red arrows). A higher (smaller) population in the qubit manifold corresponds to a larger (smaller) average interaction strength $J$.}
    \label{init}
\end{figure}

The experimental setup is shown in Fig.~\ref{setup}. For state initialization and optical readout of the ground-state spin, we address two optical transitions (labeled A and F) with a separation of $6.118$ GHz (Fig.~\ref{init}a). A 984 nm laser is locked to the frequency of the A transition. A frequency sideband at the F transition is generated by an electro-optic modulator (EOM), which is driven by radio frequency (RF) pulses with a carrier frequency of $6.118$ GHz. These RF pulses are shaped by an RF switch, controlled via a transistor-transistor logic (TTL) signal from an arbitrary waveform generator (AWG), and amplified to drive the EOM, ensuring that the optical carrier power at the A transition frequency is minimized. The optical pulses addressing the A and F transitions are shaped by two sequential acousto-optic modulators (AOMs) with up to $200$ MHz chirping range (we compromised the power to achieve a higher bandwidth, covering the optical inhomogeneous linewidth of $150$ MHz~\cite{lei2023many}). Both AOMs are driven directly by the amplified RF signal output from the AWG. The optical light is sent to the nanophotonic device in the dilution refrigerator, where Yb ions are incorporated, and the reflected light is collected by a superconducting nanowire single photon detector (SNSPD). A gating AOM is used before the SNSPD to selectively attenuate the intense reflected input pulses (the RF drive for this AOM is not shown in Fig.~\ref{setup}a). 

For microwave electronics, we need to coherently drive two spin transitions: $f_g\approx 0.675$ GHz for the ground state manifold and $f_e\approx 3.37$ GHz for the excited state manifold (Fig.~\ref{init}a). For $f_g$ ($f_e$), RF pulses with a carrier frequency of $200$ ($150$) MHz are mixed with the local oscillator (LO) frequency of $875$ ($3220$) MHz for frequency up-conversion to the target frequency. The unwanted sideband and LO components are then filtered out, and the target frequency signal is amplified. The RF signals for $f_g$ and $ f_e$ are combined using a diplexer and then sent to the device through a coplanar waveguide (Fig.~\ref{setup}b). 

Further details about the device can be found in \cite{lei2023many}.

\subsection{Experimental sequences}
The general experimental sequences are shown in Fig.~\ref{init}, including state initialization, spin dynamics control, and optical readout. Fig.~\ref{init}a shows the energy diagram of Yb, where both the ground and optically excited states exhibit fine structures labeled as \{$\ket{0}, \ket{1}, \ket{\text{Aux}}$\} and \{$\ket{0}_e, \ket{1}_e, \ket{\text{Aux}}_e$\}, respectively. As illustrated in Fig.~\ref{init}b, the sequences begin by driving the F transition for $100$~$\mu$s, during which the carrier frequency is swept over a chirping range of 200 MHz, with a repetition number of 10, to transfer population from the auxiliary state $\ket{\text{Aux}}$ to the qubit manifold. Then, a sequence involving driving on transition $A$ for $10$~$\mu$s, followed by a $\pi$ pulse resonant with the excited spin transition at frequency $f_e$, is repeated 10 times to polarize the population to $\ket{0}$. After initialization, many-body spin dynamics within the qubit manifold are interrogated using spin control sequences, such as a spin-echo sequence applied to the $f_g$ transition. At the end of the sequence, a $1$~$\mu$s optical pulse is applied to the A transition for reading out the population in $\ket{1}$.

\begin{figure}[t!]
    \includegraphics[width=\linewidth]{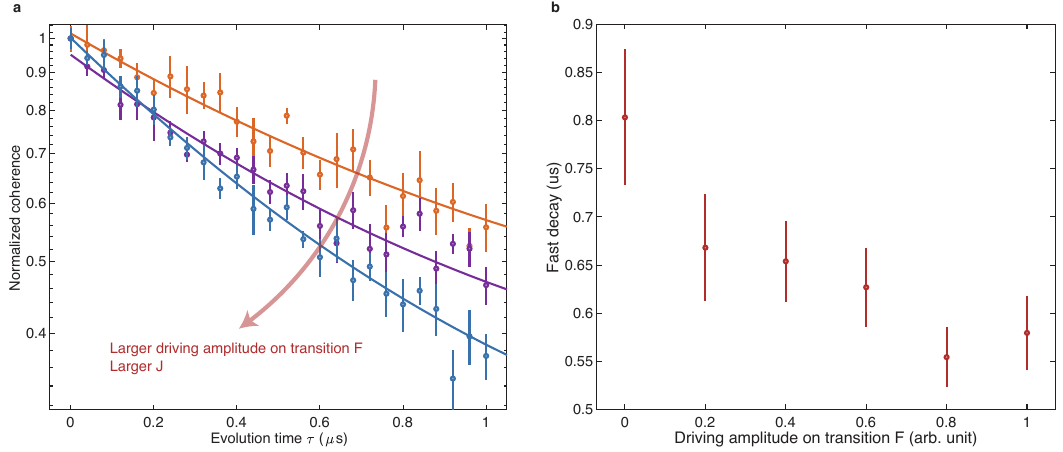}
    \caption{\textbf{Decoherence dominated by spin-spin interactions in spin-echo measurements.} \textbf{a,} Normalized coherence decay of the spin ensemble as a function of evolution time in the spin-echo sequence. The early-time decay within $1~\mu s$ accelerates with increasing $J$, achieved by increasing the driving amplitude on the F transition during state initialization. Error bars represent the standard deviation of the experimental measurements (see Sec.~\ref{cohe} for details), while the solid lines are simple exponential fits used to extract the $1/e$ decoherence times. \textbf{b,} The fitted decoherence times monotonically decrease with increasing drive amplitude on the F transition, implying that a larger population within the qubit manifold leads to stronger spin-spin interactions and thus faster decoherence.}
    \label{echomore}
\end{figure}

As shown in Fig.~\ref{init}c, we can adjust the average interaction strength $J$ between spins by tuning the population within the qubit manifold. This adjustment is achieved by varying the driving amplitude on the F transition during initialization, which controls the amount of population transferred from the auxiliary state $\ket{\text{Aux}}$ to the qubit manifold \{$\ket{0}, \ket{1}$\}. Specifically, we experimentally confirm that a higher (smaller) population in the qubit manifold leads to a larger (smaller) average interaction strength $J$, as evidenced by the spin density-dependent decoherence rates observed in the spin-echo sequence (Fig. \ref{echomore}a). The early-time decay rate within the first $1~\mu$s is governed by the average spin-spin interaction strength $J$, which exhibits a monotonic scaling with the driving amplitude on the F transition (Fig. \ref{echomore}b). This arises from the accelerated pump-out rate from the $\ket{\text{Aux}}$ state due to strong driving on transition F.

\subsection{Coherence measurement} \label{cohe}
\begin{figure}[h!]
    \centering
    \includegraphics[width=\linewidth]{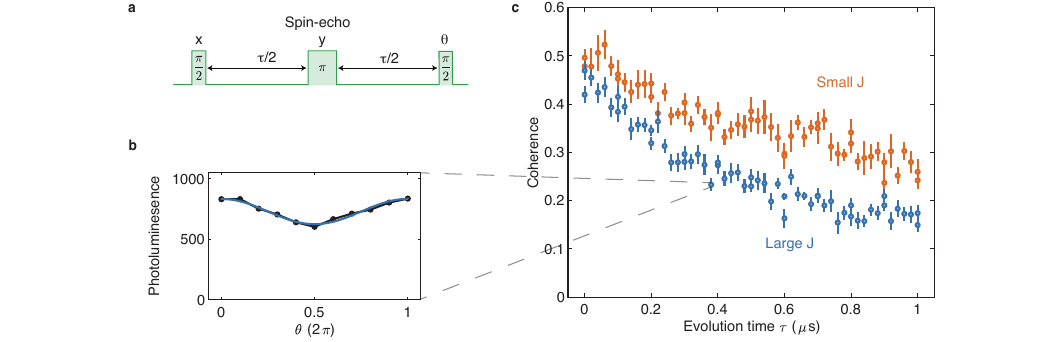}
    \caption{\textbf{Coherence characterization in spin-echo measurements.} \textbf{a,} The spin-echo sequence with a final $\pi/2$ analyzer pulse with a variable phase angle $\theta$. The phase angle $\theta$ defines the rotation axis of the $\pi/2$ pulse relative to the initial $\pi/2$ pulse. While maintaining a fixed free evolution time $\tau$, $\theta$ is varied from 0 to $2\pi$ to quantify the residual coherence of the spin ensemble. \textbf{b,} Photoluminescence signals from the spin system exhibit a sinusoidal oscillation as $\theta$ is swept, from which coherence is extracted as the contrast of the oscillation. \textbf{c,} Coherence decay profiles as a function of evolution time $\tau$ for the cases of large $J \approx 2\pi \times 0.35$ MHz and small $J \approx 2\pi \times 0.19$ MHz.}
    \label{echodetails}
\end{figure}

For all the experiments presented in the main text, the initial state is prepared as a globally polarized state along the $y$-axis. The ensemble coherence of the many-body spin system can be understood as the mean coherence of individual spins, quantified by examining the residual spin polarization along the $y$-axis at the end of spin dynamics interrogation (Fig.~\ref{echodetails}a). In other words, measuring coherence is equivalent to measuring the polarization along the $y$-axis. Note that the $\pi/2$ analyzer pulse converts the $y$-axis polarization into $z$-axis polarization, which can be read out using photoluminescence signals $C$. By sweeping the phase angle, $\theta$, of the analyzer pulse (at a fixed interrogation time), we obtain sinusoidal oscillations of photoluminescence as a function of $\theta$ and fit them with $C(\theta) = C_\text{amp} \cos{\theta} + C_\text{offset}$ (Fig.~\ref{echodetails}b). The coherence, $\tilde{C}$, is then calculated as $\tilde{C} = \frac{C_\text{max} - C_\text{min}}{C_\text{max} + C_\text{min}} = \frac{C_\text{amp}}{C_\text{offset}}$, and the corresponding error bar is extracted from the self-covariance of the fit (Fig.~\ref{echodetails}c).

\subsection{Microscopic numerical simulation}\label{numsimu}
\begin{figure}
    \centering \includegraphics[width=0.95\linewidth]{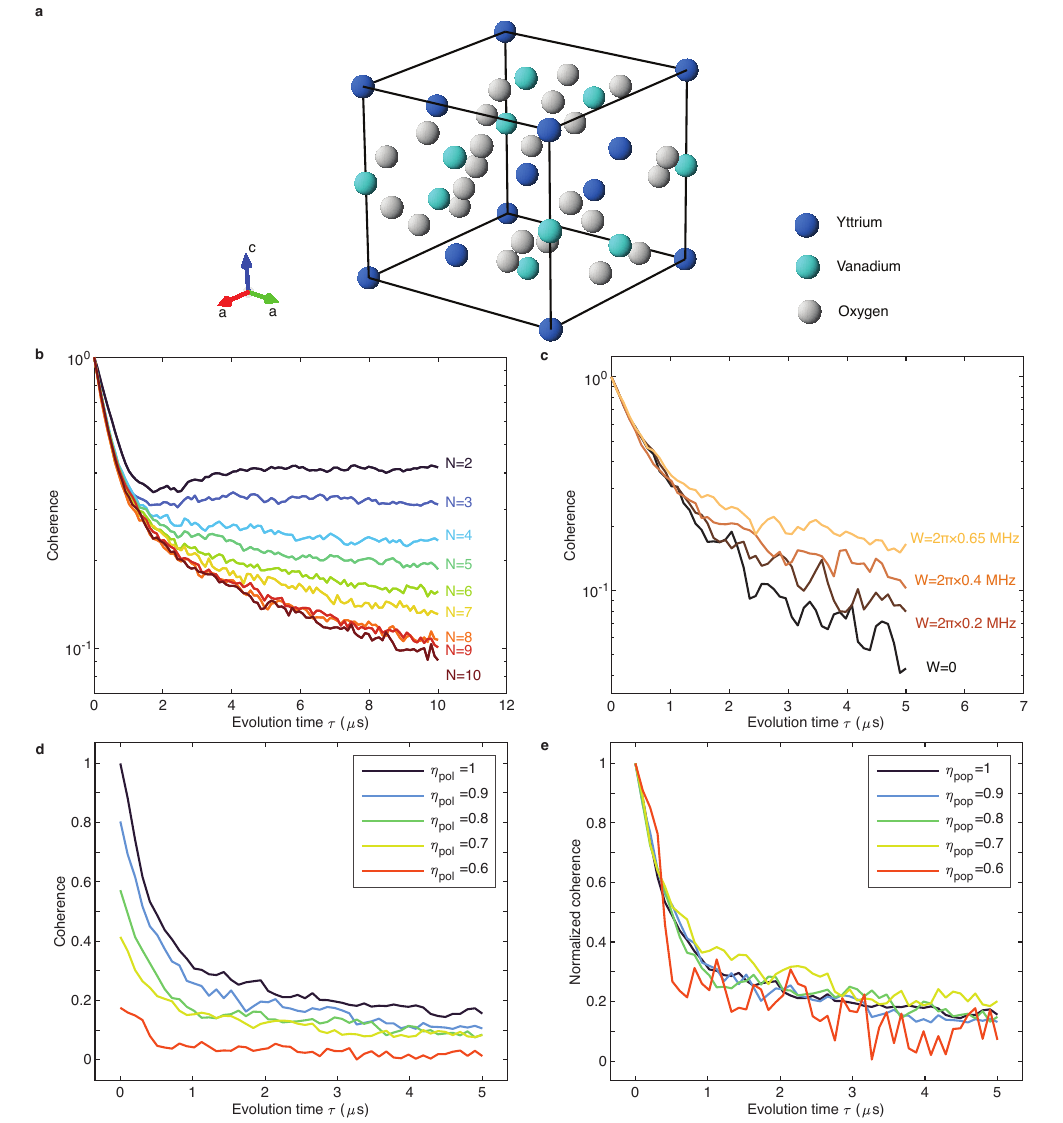}
    \caption{\textbf{Numerical simulation results of the spin-echo decoherence dynamics.} \textbf{a,} The crystal structure of YVO$_4$. \textbf{b,} System size scaling of the spin-echo decoherence profiles for different numbers of simulated spins from $N=2$ to $N=10$ ions. The decay profiles start to converge approximately when $N > 8$ ions. \textbf{c,} Effects of on-site disorder on spin-echo decoherence dynamics, simulated for different disorder strengths $W$. After the early-time transient ($ t < 1~\mu$s) where the decay rate is independent of on-site disorder, spin decoherence slows down, indicating a crossover from an interaction-dominated to a disorder-dominated regime. \textbf{d,} Effects of initial spin polarization, $\eta_\text{pol}$, on spin-echo decoherence dynamics. \textbf{e,} Normalized coherence for the decay time traces shown in \textbf{d}. Upon rescaling, the decay profiles overlap for different polarization values of $\eta_\text{pol}$, indicating that the decoherence dynamics are independent of initial spin polarization.}
    \label{MB_simu_fig}
\end{figure}

Here, we present the details of the numerical simulations used throughout this work. To set up the spin system in simulations, we begin by generating the lattice structure of YVO$_4$ (Fig.~\ref{MB_simu_fig}a). We center the lattice on a Y ion and replace it with a Yb ion, then the lattice size is appropriately extended to accommodate the doping concentration of Yb ions for $N$ ions. These $N$ ions are randomly substituted into the Y sites (including the central one) with random spin frequency detuning $\Delta_i$. The values of $\Delta_i$ are randomly sampled from a Lorentzian distribution with a full width at half maximum $W$. This defines the on-site disorder Hamiltonian, $\hat{H}_\text{dis}$, given by 
\begin{equation}
    \hat{H}_\text{dis} = \sum_{i=1}^N \Delta_i \hat{S}_z^i.
\end{equation}
Regarding the modeling of spin-spin interactions, the pairwise interaction strength $J_{ij}$ between each pair of Yb ions is calculated based on their dipolar XY interaction Hamiltonian, $\hat{H}_\text{int}$:
\begin{align} \label{Hdd_SI}
    \hat{H}_\text{int}&= \sum_{ij,\; i>j}^N J_{ij} (\hat{S}_x^i\hat{S}_x^j+\hat{S}_y^i\hat{S}_y^j), \\
    J_{ij} &= -\frac{\mu_0\mu_B^2g_{\parallel}^2}{8\pi r_{ij}^3}(3z_{ij}^2-1). \label{Jij}
\end{align}
Here, $\mu_0$ is the magnetic permeability of free space, $g_{\parallel}=6.08$ is the g-factor along the crystal $c$-axis~\cite{Kindem2018} (defining the $c$-axis to be the $z$-axis), $r_{ij}$ is the distance between  ions $i$ and $j$, and $z_{ij} \equiv |\vec{r}_{ij} \cdot \vec{e}_z| / r_{ij}$ is the $z$-directional cosine between the ions ($-1$$\le$$z_{ij}$$\le$$1$), where $\vec{e}_z$ is the unit vector along the $z$-axis. Since $J_{ij}$ is dependent on both distance and orientation between each spin pair, we define the average spin-spin interaction $J$ as follows:
\begin{equation}
    J \equiv \frac{\mu_0\mu_B^2g_{\parallel}^2}{4\pi} \frac{1}{\left\langle r_{ij}\right\rangle^3}    = 2\pi \times 480~\text{MHz $\cdot$ nm$^3$} \times \frac{1}{\left\langle r_{ij}\right\rangle^3},
\end{equation}
where $\langle r_{ij} \rangle$ is the average nearest-neighbor distance in units of nm, and we have used the condition $z_{ij} = \pm 1$ to maximize the interaction strength in different orientations. For a given crystal structure, the average interaction strength is linearly proportional to the spin density $n_s$, i.e., $J \propto n_s$. For YVO$_4$ crystal, the lattice parameters are $a = 7.119 \textup{\AA}, c = 6.290 \textup{\AA}$. Considering that there are 4 ions per unit cell, the relation between the average distance, $\langle r_{ij} \rangle$ in units of nm, and the spin concentration, $n_s$ in units of ppm, can be expressed as $\langle r_{ij} \rangle = \frac{0.629}{(4n_s)^{1/3} \times 10^{-2}}$ nm. Since the g-factor of a Yb ion is three times larger than that of an electron, we expect the interaction strength in our system to be approximately 9 times larger compared to the case of an NV center, where $J = 2\pi \times 52~\text{MHz $\cdot$ nm$^3$} \times \frac{1}{\left\langle r_{ij}\right\rangle^3}$~\cite{davis2023probing}. 

Given a random realization of a disordered spin system composed of $N$ Yb ions, the many-body spin states can be fully characterized by solving for noiseless dynamics under the system Hamiltonian $\hat{H} = \hat{H}_\text{dis} + \hat{H}_\text{int}$, incorporating control pulse sequences such as spin-echo or Ramsey sequences. To mitigate numerical artifacts stemming from boundary effects in finite-size simulations, only the final state of the central Yb ion is considered for readout. This process is repeated across multiple Monte Carlo runs, where the positions and on-site detunings of the ions are randomized each time, to capture ensemble-averaged dynamics.

To determine the number of ions needed to simulate the system accurately, a convergence test on system size $N$ is conducted (Fig.~\ref{MB_simu_fig}b). We find that the dynamics start to converge for system sizes larger than 8 ions, as adding ions farther away does not significantly contribute to the dynamics of the central readout spin. Henceforth, we fix the system size to $N=9$ ions in Fig.~\ref{MB_simu_fig}c-e, as well as the simulation results presented in the main text.

\begin{figure}
    \centering    \includegraphics[width=\linewidth]{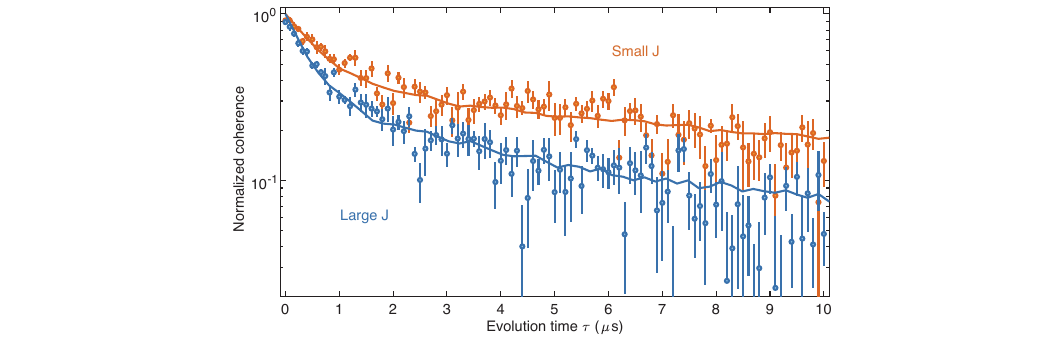}
    \caption{\textbf{Comparison of long-time decoherence dynamics between experiment and simulation under the spin-echo dynamics.} The numerical simulations with calibrated parameters (solid lines) show excellent agreement with the experiments (markers) over extended long times for both the small $J$ (red) and large $J$ (blue) cases.}
    \label{echo_longer}
\end{figure}

Based on the many-body simulation of a disordered spin ensemble, we extract the average interaction strength, $J$, and on-site disorder, $W$, by comparing the simulation results to the experimentally observed dynamics. First, the early-time decay rate within $1~\mu s$ does not depend on on-site disorder strength $W$; it is only governed by the spin-spin interaction $J$ (Fig.~\ref{echomore}c). This allows us to determine the average interaction strength, where the only unknown parameter is $n_s$, the concentration of Yb ions. By matching the decay rate at short times ($t < 1~\mu s$) from the simulation to that of the experiment, we can extract the ion concentration within the qubit manifolds for different cases, yielding 25 ppm for the small $J$ case and 46 ppm for the large $J$ case (see Fig.~2a in the main text). The extracted effective spin densities within the qubit manifold are estimated to be lower than an independently measured concentration of 86 ppm using secondary ion mass spectrometry, which reflects the {\it total} density of Yb ions~\cite{bartholomew2020chip}. This difference is attributed to the remaining untransferred population residing in the doubly degenerate $\ket{\text{Aux}}$ states outside the qubit manifold. We then extract $W = 2\pi \times 0.65$~MHz,  independent of the interaction strength $J$, by finding the best agreement in the experiment and simulation under the Ramsey sequence (see Fig.~2b in the main text). As shown in Fig.~\ref{MB_simu_fig}c, the presence of on-site disorder slows down the decoherence dynamics at late times in the spin-echo sequence. We successfully match this extended-time behavior using the disorder strength $W$, extracted from Ramsey data (Fig.~\ref{echo_longer}). 

These calibrated interaction and disorder strengths reproduced all the experimental data across all cases presented in the main text figures. This strongly suggests that the decoherence dynamics of the spin ensemble are driven by the intrinsic spin-spin interactions and on-site disorder described by the {\it closed} many-body Hamiltonian, without additional dephasing due to extrinsic coupling to external environments.

We also investigate whether the initial spin distribution within the qubit manifold influences the resulting decoherence processes. Specifically, as described earlier, we employ both optical transition $A$ and microwave transition $f_e$ to initialize the Yb ion's population into the $\ket{0}$ state. We can quantify the fidelity of this process as initial polarized rate $\eta_\text{pol} \equiv \frac{p_0}{p_0 + p_1}$, where $p_0$ and $p_1$ correspond to the populations in $\ket{0}$ and $\ket{1}$ after initialization, respectively. The relation between $\eta_\text{pol}$ and the measured coherence $\tilde{C}$ (Sec.~\ref{cohe}) is $\tilde{C}=2\eta_\text{pol}-1$. According to the experimental data shown in Fig.~\ref{echodetails}c near the $\tau=0$ point, we estimate that $\eta_\text{pol} \approx 0.75$. From numerical simulations investigating the dependence of decoherence dynamics on initial polarized rates, we find that the decoherence profiles do not depend on the value of $\eta_\text{pol}$, except for a reduction in the contrast of the coherence signal due to imperfect initialization (Fig.~\ref{MB_simu_fig}d). We confirm that all cases overlap when each imperfect case is rescaled by a factor of $1/(2\eta_\text{pol} - 1)$ or equivalently by the maximum contrast at $\tau=0$ (Fig.~\ref{MB_simu_fig}e). The independence of the initial polarized rate arises because, despite individual spins being randomly oriented along either the $+y$ or $-y$ directions, the dipolar XY Hamiltonian (Eq.~\ref{Hdd_SI}) induces dephasing dynamics through terms like $\hat{S}^i_x \hat{S}^j_x$, which flip ions $i$ and $j$, regardless of whether they are initially aligned parallel or antiparallel along the $y$-axis.

\section{Theoretical modeling and analysis}
\subsection{Derivation of the dipolar XY Hamiltonian} \label{SecXY}
Here, we provide a theoretical analysis for constructing the dipolar XY Hamiltonian in our system (Eq.~\ref{Hdd_SI}). We start with the magnetic dipole-dipole interaction and derive an effective spin-spin interaction within the qubit manifold. The ground-state qubit states \{$\ket{0}$,$\ket{1}$\} at zero magnetic field can be expressed as:
\begin{equation}
    \ket{0}=\frac{1}{\sqrt{2}}(\ket{\uparrow\Downarrow}-\ket{\downarrow\Uparrow})
\end{equation}
\begin{equation}
    \ket{1}=\frac{1}{\sqrt{2}}(\ket{\uparrow\Downarrow}+\ket{\downarrow\Uparrow})
\end{equation}
where $\{\ket{\uparrow},\ket{\downarrow}\}$ are electron spins and $\{\ket{\Uparrow},\ket{\Downarrow}\}$ are nuclear spins associated with a Yb ion~\cite{Kindem2018}. The magnetic dipole-dipole interaction Hamiltonian between the spins of ions $i$ and $j$ is described as follows:
\begin{equation} \label{Hddinit}
    \hat{H}_{dd}^{ij}=-\frac{\mu_0}{4\pi r_{ij}^3}(3(\vec{\mu}_i\cdot\vec{e}_{r_{ij}})(\vec{\mu}_j\cdot\vec{e}_{r_{ij}})-\vec{\mu}_i\cdot\vec{\mu}_j)
\end{equation}
where $\vec{e}_{r_{ij}} \equiv \vec{r}_{ij}/r_{ij}$ is the normalized unit distance vector from ion $i$ to ion $j$, and $\vec{\mu}_j$ is the magnetic dipole moment operator for ion $j$, given as 
\begin{equation}
    \vec{\mu}_j = \mu_B\textbf{g}\cdot{\textbf{S}_{0}^j} = \mu_B\begin{pmatrix}
g_{\bot}\hat{S}_{0,x}^j  \\
g_{\bot}\hat{S}_{0,y}^j  \\
g_{\parallel}\hat{S}_{0,z}^j 
\end{pmatrix}
\end{equation}
where $\mu_B$ is the Bohr magneton, $\textbf{g} = \text{diag}(g_{\bot}, g_{\bot}, g_\parallel)$ is the anisotropic g-factor tensor with $g_\bot = 0.85$ and $g_\parallel = 6.08$, and $\textbf{S}_0^j=(\hat{S}_{0,x}^j, \hat{S}_{0,y}^j, \hat{S}_{0,z}^j)^T$ is the vectorized spin operator defined in Yb's electron spin basis  $\{\ket{\uparrow},\ket{\downarrow}\}$ (which is different from the $\{\ket{0},\ket{1}\}$ basis).

Since we are interested in deriving the effective interaction Hamiltonian within the qubit manifold, we first calculate the matrix elements of the magnetic dipole moment with respect to the qubit $\{\ket{0},\ket{1}\}$ basis states:

\begin{align*} \nonumber
    \bra{1}_j\vec{\mu}_j\ket{0}_j & = \bra{0}_j\vec{\mu}_j\ket{1}_j =\mu_B\begin{pmatrix}
g_{\bot}\bra{0}_j \hat{S}_{0,x}^j\ket{1}_j  \\
g_{\bot}\bra{0}_j \hat{S}_{0,y}^j\ket{1}_j  \\
g_{\parallel} \bra{0}_j \hat{S}_{0,z}^j\ket{1}_j
\end{pmatrix}
=\frac{1}{2}\mu_B\begin{pmatrix}
0 \\
0  \\
g_{\parallel}
\end{pmatrix} \\
\bra{0}_j\vec{\mu}_j\ket{0}_j &= \bra{1}_j\vec{\mu}_j\ket{1}_j = 0.
\end{align*}
Note that both $\ket{0}$ and $\ket{1}$ states have no magnetic moments. This indicates that the qubit states are first-order insensitive to external magnetic field, defining our ``clock'' transition with a zero first-order Zeeman (ZEFOZ) shift.

We then consider the matrix elements of the dipolar interaction Hamiltonian (Eq.~\ref{Hddinit}) in a two-qubit basis, $\{\ket{00}, \ket{01}, \ket{10}, \ket{11}\}$, and find that the ZEFOZ nature of our qubit leads to the following form:
\begin{equation} \label{dipole_orig}
    \hat{H}_{dd}^{ij} = \frac{J_{ij}}{2}(\hat{S}_+^i \hat{S}_-^j+\hat{S}_-^i \hat{S}_+^j+ \hat{S}_+^i \hat{S}_+^j+ \hat{S}_-^i \hat{S}_-^j).
\end{equation}
Here, $\hat{S}^j_\pm = \hat{S}^j_x\pm i\hat{S}^j_y$ are the raising and lowering spin operators for ion $j$, defined in the qubit $\{\ket{0},\ket{1}\}$ basis, and $J_{ij}$ is defined in Eq.~\ref{Jij}. The dipole-dipole interaction described above includes flip-flop, flop-flip, flip-flip, and flop-flop terms. However, the last two terms can be neglected due to energetic suppression, as the qubit transition frequency, $f_g \approx 675$ MHz, is much larger than the interaction strength, $J_{ij}/2\pi \sim 0.1$ MHz. This approximation is known as the secular approximation, which leads to the following pure spin-exchange pairwise interaction:
\begin{align} \label{Hdd}
    \hat{H}_{dd}^{ij} \approx \frac{J_{ij}}{2}(\hat{S}_+^i \hat{S}_-^j+\hat{S}_-^i \hat{S}_+^j) = J_{ij}( \hat{S}_x^i \hat{S}_x^j+ \hat{S}_y^i \hat{S}_y^j),
\end{align}
consistent with the interaction Hamiltonian between ions $i$ and $j$, as presented in Eq.~\ref{Hdd_SI}.

\subsection{Tunable dipolar XXZ Hamiltonians}
In the absence of an external magnetic field, our system Hamiltonian is described by the dipolar XY model, which contains only spin-exchange interactions (Eq.~\ref{Hdd_SI}). However, applying a small magnetic field along the $z$-axis allows us to achieve a tunable spin model that includes Ising interactions, known as the XXZ model. Specifically, with a small magnetic field strength $B$, the original eigenstates $\ket{0}$ and $\ket{1}$ are perturbed as follows~\cite{ruskuc2022nuclear}:
\begin{align}
    \ket{\tilde{0}}&=\ket{0}-\frac{\alpha_B}{2} \ket{1} \\
     \ket{\tilde{1}}&=\ket{1}+\frac{\alpha_B}{2} \ket{0}
\end{align}
Here, $\{\ket{\tilde{0}}, \ket{\tilde{1}}\}$ are the perturbed eigenstates with an admixture ratio $\alpha_B \equiv g_\parallel \mu_B B / \omega$, and $\omega = 2\pi f_g = 2\pi\times 675$ MHz is the angular qubit transition frequency. Under these perturbed eigenstates, the system's original Hamiltonian, shown in Eq.~\ref{Hdd_SI}, can be modified as
\begin{align} \label{newHdd}
    \hat{H}_\text{int}&= \left( 1- \frac{\alpha_B^2}{2} \right) \hat{H}_\text{exchange}+ 2\alpha_B^2 \hat{H}^z_\text{Ising}
\end{align}
where the spin-exchange Hamiltonian, $\hat{H}_\text{exchange}$, and the Ising Hamiltonian, $\hat{H}^z_\text{Ising}$ are given by
\begin{align}
    \hat{H}_\text{exchange} &=  \sum_{ij, \; i>j}^N J_{ij} (\tilde{S}_x^i\tilde{S}_x^j+\tilde{S}_y^i\tilde{S}_y^j) \\
    \hat{H}^z_\text{Ising} &= \sum_{ij, \; i>j}^N J_{ij} \tilde{S}_z^i \tilde{S}_z^j.
\end{align}
We have used the spin operators for the new perturbed basis, $\{\ket{\tilde{0}}, \ket{\tilde{1}}\}$, and applied the secular approximation to preserve the energy-conserving term. From the new Hamiltonian in Eq.~\ref{newHdd}, one can observe that applying an external magnetic field tunes the magnetic field-dependent admixture ratio, $\alpha_B$, thereby controlling the relative strengths of the spin-exchange interaction, $\hat{H}_\text{exchange}$, and the Ising interaction $\hat{H}^z_\text{Ising}$. 

\subsection{Analysis of influences from the $\ket{\text{Aux}}$ states}
In the main text, we consider only the effective interactions within the qubit manifold, despite the presence of the doubly-degenerate $\ket{\text{Aux}}$ state in the ground state. Using the electron and nuclear spin notations introduced earlier, the two degenerate states are expressed as $\ket{\uparrow\Uparrow}$ and $\ket{\downarrow\Downarrow}$. The populations in these $\ket{\text{Aux}}$ states can also interact with the spins in the qubit manifold. In this section, we will analyze the influence of the $\ket{\text{Aux}}$ state on the qubit manifold.

Specifically, the auxiliary states can interact with the qubit states via the full dipole-dipole Hamiltonian given in Eq.~\ref{Hddinit}. Following a similar procedure presented in Sec.~\ref{SecXY}, we first calculate the matrix element of the magnetic moment operator between $\ket{\text{Aux}}$ and one of the qubit states. For instance, between $\ket{\uparrow\Uparrow}$ and $\ket{0}$, we have:
\begin{equation}
    \bra{0}_j\vec{\mu}_j\ket{\uparrow\Uparrow}_j=\mu_B\begin{pmatrix}
g_{\bot}\bra{0}_j \hat{S}_{0,x}^j\ket{\uparrow\Uparrow}_j  \\
g_{\bot}\bra{0}_j \hat{S}_{0,y}^j\ket{\uparrow\Uparrow}_j  \\
g_{\parallel} \bra{0}_j \hat{S}_{0,z}^j\ket{\uparrow\Uparrow}_j
\end{pmatrix}
=\frac{g_{\bot}}{2\sqrt{2}}\mu_B\begin{pmatrix}
-1 \\
i  \\
0
\end{pmatrix}
\end{equation}
Next, we consider the spin-exchange process between ion $i$ at $\ket{\text{Aux}}$ and ion $j$ at $\ket{0}$, such as $\ket{\uparrow\Uparrow}_i \ket{0}_j \leftrightarrow \ket{0}_i \ket{\uparrow\Uparrow}_j$. The corresponding relevant matrix element of the Hamiltonian can be computed as
\begin{equation}
\bra{0}_i\bra{\uparrow\Uparrow}_j\hat{H}_{dd}^{ij}\ket{\uparrow\Uparrow}_i\ket{0}_j=\frac{\mu_0\mu_B^2g_{\bot}^2}{32\pi r_{ij}^3}(3z_{ij}^2-1) = -\frac{1}{4} \left(\frac{g_{\bot}}{g_{\parallel}}\right)^2 J_{ij}
\end{equation}
where $J_{ij}$ is the spin-exchange interaction strength within the qubit manifold defined in Eq.~\ref{Jij}. Note that the spin-exchange interaction strength between $\ket{\text{Aux}}$ and the qubit states is significantly smaller by a factor of $(g_{\bot} / g_{\parallel})^2 \approx 0.02$. Consequently, using Fermi's golden rule, the transition rate from $\ket{0}$ to $\ket{\text{Aux}}$ can be compared against that from $\ket{0}$ to $\ket{1}$:
\begin{equation}
    \frac{\Gamma_{\ket{0}\rightarrow \ket{\text{Aux}}}}{\Gamma_{\ket{0}\rightarrow \ket{1}}} = \frac{|\bra{0, \text{Aux}}\hat{H}^{ij}_{dd} \ket{\text{Aux}, 0}|^2}{|\bra{0,1} \hat{H}^{ij}_{dd} \ket{1,0}|^2} \sim \left(\frac{g_{\bot}}{g_{\parallel}}\right)^4 \sim 10^{-4}.
\end{equation}
This implies that the population exchange between the ions in the $\ket{\text{Aux}}$ state and the ions in the qubit manifold occurs at a rate that is approximately $10^4$ times slower than the population exchange within the qubit manifold. Additionally, the larger on-site disorder of the $\ket{\text{Aux}}$ states, due to their sensitivity to external fields, further prevents this flip-flop process between $\ket{\text{Aux}}$ and the qubit manifold, ruling out its contribution to the late-time slow decoherence observed in the spin-echo experiment.

\section{Detailed analysis for $\epsilon$-CPMG sequence}
\begin{figure}[hbt!]
    \centering
    \includegraphics[width=\linewidth]
    {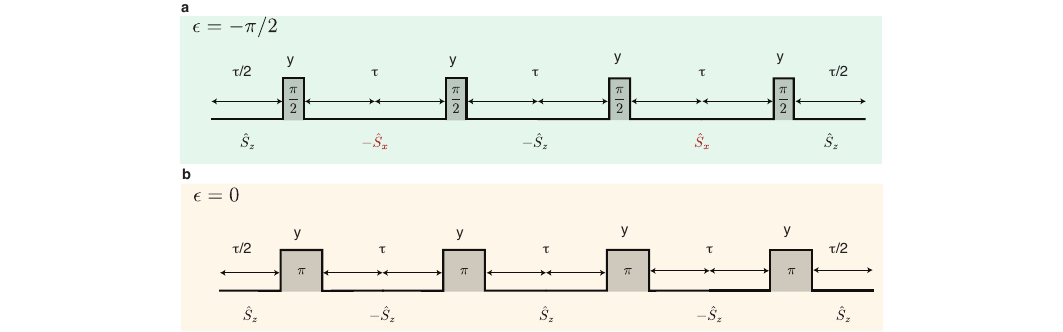}
    \caption{\textbf{Interaction-picture-based toggling-frame transformation of the $\hat{S}_z$ operator for $\epsilon$-CPMG sequences.} \textbf{a,} $\epsilon = -\pi/2$ is a preferable choice for decoupling spin-spin interactions in a strongly interacting spin system. \textbf{b,} $\epsilon = 0$ corresponds to the conventional CPMG sequence, which decouples on-site disorder through periodic $\pi$ pulses. These base sequences are repeated in time stroboscopically with fixed pulse spacing $\tau$ in dynamic Hamiltonian engineering.}
    \label{CPMGdetail}
\end{figure}

The $\epsilon$-CPMG sequence is employed to verify the presence of strong spin-spin interactions in our system due to its distinct sensitivity to interaction and on-site disorder (see Fig. 2 of the main text). When $\epsilon=0$, it is identical to the standard CPMG sequence, effectively decoupling time-independent on-site disorder while leaving pairwise spin-spin interactions unchanged (Fig.~\ref{CPMGdetail}b). In contrast, when $\epsilon=\pm\pi/2$, the initial spin states polarized along the $y$-axis can be protected against both on-site disorder and spin-spin interactions to leading order. This is consistent with experimental observations showing that coherence at $\epsilon=\pm \pi/2$ is higher than at $\epsilon=0$ (see Fig.~2c of the main text). 

In the following, we will derive the zeroth-order and first-order Hamiltonians for $\epsilon$-CPMG sequences at $\epsilon = 0$ and $-\pi /2$ via average Hamiltonian theory~\cite{choi2020robust}, starting from the parent system Hamiltonian $\hat{H}$:
\begin{equation} \label{H}
    \hat{H} = \hat{H}_\text{dis} + \hat{H}_\text{int} =\sum_{i=1}^N\Delta_i \hat{S}_z^i+\sum_{ij, \; i>j }^N J_{ij}(\hat{S}_x^i \hat{S}_x^j+\hat{S}_y^i \hat{S}_y^j)
\end{equation}
The effective average Hamiltonians, $\hat{H}_\text{eff}$, dynamically engineered from $\hat{H}$, can be analytically computed order-by-order using the framework based on the interaction-picture-based toggling-frame transformation of the $\hat{S}_z$ operator~\cite{choi2020robust}. Specifically, the system Hamiltonian $\hat{H}$ transforms into an effective Hamiltonian $\hat{H}_\text{eff}$ as follows:
\begin{equation} \label{Heff}
    \hat{H} \rightarrow \hat{H}_\text{eff} = \sum_{m=0}^\infty \hat{H}_m
\end{equation}
where $\hat{H}_m$ denotes the average Hamiltonian of order $m$ determined by a control pulse sequence.

\subsection{$\epsilon=-\pi/2$}
Here we consider the $-\frac{\pi}{2}$-CPMG sequence with the base control sequence depicted in Fig.~\ref{CPMGdetail}a. 
The zeroth-order Hamiltonian, $\hat{H}_0$, is given as
\begin{equation}
    \hat{H}_0 = \frac{1}{2} \sum_{ij, \; i>j}^N J_{ij} (\hat{S}_y^i \hat{S}_y^j+\vec{S^i} \cdot\vec{S^j}) = \frac{1}{2} \hat{H}^y_\text{Ising} + \frac{1}{2} \hat{H}_\text{Heis},
\end{equation}
where $\hat{H}^y_\text{Ising} = \sum_{ij, \; i>j}^N J_{ij} \hat{S}_y^i \hat{S}_y^j$ is the Ising interaction along the $y$-axis, and $\hat{H}_\text{Heis} = \sum_{ij, \; i>j}^N J_{ij}\vec{S^i}\cdot\vec{S^j}$ is the Heisenberg interaction. Note that the initially polarized spin state along the $y$-axis is the eigenstate of $\hat{H}_0$, resulting in the preservation of coherence.

The next first-order Hamiltonian, $\hat{H}_1$, is given as
\begin{equation}
    \hat{H}_1 =\frac{\tau}{4}\left[\sum_i^N \Delta_i^2 \hat{S}_y^i + \sum_{ij, \; i\neq j}^N \Delta_i J_{ij}\left(2 \hat{S}_z^i \hat{S}_y^j-\hat{S}_y^i \hat{S}_z^j\right) \right],
\end{equation}
which gives rise to the dephasing of the polarized spin ensemble. In principle, symmetrizing the pulse sequence can eliminate all odd-order average Hamiltonians~\cite{choi2020robust}, potentially extending the coherence time. We leave it for future work.

\subsection{$\epsilon=0$}
Here we consider the $\epsilon=0$ case corresponding to the conventional CPMG sequence with the base control sequence depicted in Fig.~\ref{CPMGdetail}b. Its zeroth-order Hamiltonian, $\hat{H}_0$, is given as
\begin{equation}
    \hat{H}_0 =\sum_{ij, \; i>j}^N J_{ij}(\hat{S}_x^i \hat{S}_x^j+\hat{S}_y^i \hat{S}_y^j) = \hat{H}_\text{Heis} - \hat{H}^z_\text{Ising}
\end{equation}
where $\hat{H}^z_\text{Ising}$ is the Ising interaction along the $z$-axis. The initially polarized spin state along the $y$-axis undergoes dephasing with $\hat{H}_0$ because the $\hat{H}^z_\text{Ising}$ term provides an interaction-induced effective magnetic field along the $z$-axis.

The next first-order Hamiltonian, $\hat{H}_1=0$. In fact, all odd-order average Hamiltonians of the CPMG sequence are zero due to the mirror symmetry of the toggling-frame transformations~\cite{choi2020robust}.

\subsection{Comparsion between $\epsilon=-\pi/2$ and $\epsilon=0$ cases}
\begin{figure}[h!]
    \includegraphics[width=\linewidth]{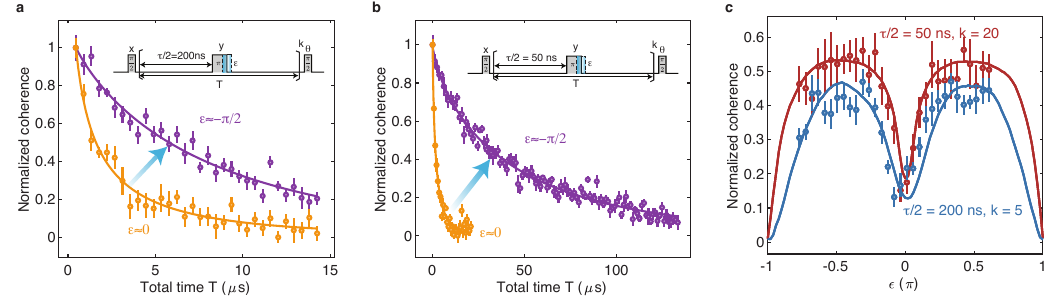}
    \caption{\textbf{Coherence characterization in $\epsilon$-CPMG measurements.} \textbf{a, b,} We compare coherence decay profiles as a function of total interrogation time $T$ for two different $\epsilon$-CPMG measurements at $\epsilon \approx 0$ (yellow) and $\epsilon \approx -\pi/2$ (purple). For both cases, the corresponding base sequence has a periodicity of $\tau + t_p$, where $\tau$ is the free evolution period and $t_p$ is the pulse duration, and is repeated $k$ times to advance in time (see the insets). We measure coherence at stroboscopic times $T = k (\tau + t_p)$. \textbf{a,} $\tau/2 = 200$~ns, $t_p = 45$~ns for $\epsilon \approx 0$; $t_p = 25$~ns for $\epsilon \approx -\pi/2$. \textbf{b,} $\tau/2 = 50$~ns, $t_p = 43$~ns for $\epsilon \approx 0$; $t_p = 22$~ns for $\epsilon \approx -\pi/2$. Experimental data are obtained under the large $J$ condition, and solid lines represent phenomenological stretched exponential fits. To facilitate comparison between the two different $\epsilon$ cases, the coherence decay profiles are normalized by their respective maximum coherence values. \textbf{c,} Dependence of $\epsilon$-CPMG sequence coherence on $\epsilon$ for the long $\tau$ (blue) and short $\tau$ (red) cases. When sweeping $\epsilon$, fixed base sequence parameters of $(\tau/2 = 200~\text{ns}, k=5;~\text{blue})$ and $(\tau/2 = 50~\text{ns}, k=20;~\text{red})$ are used respectively. Here, all experimental data across different $\epsilon$ values are globally normalized by the initial state polarization, $\eta_\text{pol}$~(see Sec.~\ref{numsimu} and Fig.~\ref{MB_simu_fig}d,e). The solid lines denote numerical simulation results using the experimentally calibrated system parameters. The simulated results were globally rescaled to facilitate comparison with the experimental data.} 
    \label{CPMG_time}
\end{figure}

The time traces of spin ensemble coherence for the $\epsilon$-CPMG sequences at $\epsilon=0$ and $\epsilon = -\pi/2$ are shown in Fig.~\ref{CPMG_time}a,b. Consistent with the average Hamiltonian analysis, the $\epsilon$-CPMG sequence with $\epsilon=-\pi/2$ exhibits a significant enhancement in coherence, attributed to the zeroth-order Hamiltonian supporting the initially polarized state as an eigenstate. It is noteworthy that the rapid driving of the base sequence with a shorter periodicity leads to a much longer coherence time, which corroborates the effectiveness of the zeroth-order Hamiltonian in dominating the engineered many-body dynamics in the fast driving regime. Furthermore, when the rotation angle offset, $\epsilon$, is swept at a fixed sequence repetition number $k$, the $\epsilon$-CPMG sequence reveals an unconventional profile (Fig.~\ref{CPMG_time}c). This profile is in stark contrast to the behavior expected in a non-interacting spin system, providing further evidence of the strong spin-spin interactions within our spin system (see Fig.~2c of the main text).

\section{Models I and II for the spin-echo measurement}
In this section, we provide a more detailed explanation of Models I and II mentioned in Fig.~3 of the main text.

In Model I, we simulate the spin-echo signal governed by only a single pair of resonant spins with the largest pairwise interaction strength in a given realization, denoted as $J_\text{max}$. The spin-exchange interaction induces periodic entanglement and disentanglement dynamics between this pair, resulting in coherence oscillations at a rate of $J_\text{max}/2$. Concretely, after the initial $\pi/2$ pulse along the $x$-axis, the quantum state is prepared as $\ket{\psi(0)} = \frac{1}{2}(\ket{0} -i \ket{1}) \otimes (\ket{0} -i \ket{1})$, which then evolves as:
\begin{align*}
    \ket{\psi(\tau)} = \frac{1}{2}\left[\ket{00} - \ket{11} -i e^{-i J_\text{max} \tau/2} (\ket{01} + \ket{10})\right]
\end{align*}
under the two-spin Hamiltonian $\hat{H} = \frac{J_\text{max}}{2}(\hat{S}_+ \hat{S}_-+ \hat{S}_- \hat{S}_+)$. As a result, the corresponding coherence oscillates sinusoidally as $\cos(J_\text{max}\tau/2)$. Note that $J_\text{max}$ is a {\it random} number determined by a stochastic realization of spin positions in the numerical simulation. Through repeated Monte Carlo sampling of different spin configurations, we calculate the ensemble average of coherence as a function of time, revealing rapid decay. Despite the simplicity of the model, we find that the early-time decay rate of Model I is consistent with that of the full simulation carried out in Model III.

In Model II, we expand upon Model I and simulate many-body dynamics involving more than two spins without disorder. Each spin can now become entangled with multiple neighboring spins simultaneously, undergoing intricate dephasing dynamics. In each simulation run, we calculate the coherence of the center spin and perform Monte Carlo averaging over different realizations of spin configurations. As seen in Fig.~3 of the main text, it is observed that adding more spins slows down the late-time decay compared to Model I, thereby improving the simulation's agreement with the experimental data.

To gain a better understanding of why Model II has slower dynamics compared to Model I at later times, we will derive and discuss two toy models---the two-spin and three-spin cases---in the following subsections.

\subsection{Two-spin case}
\begin{figure}[hbt!]
    \centering
    \includegraphics[width=\linewidth]{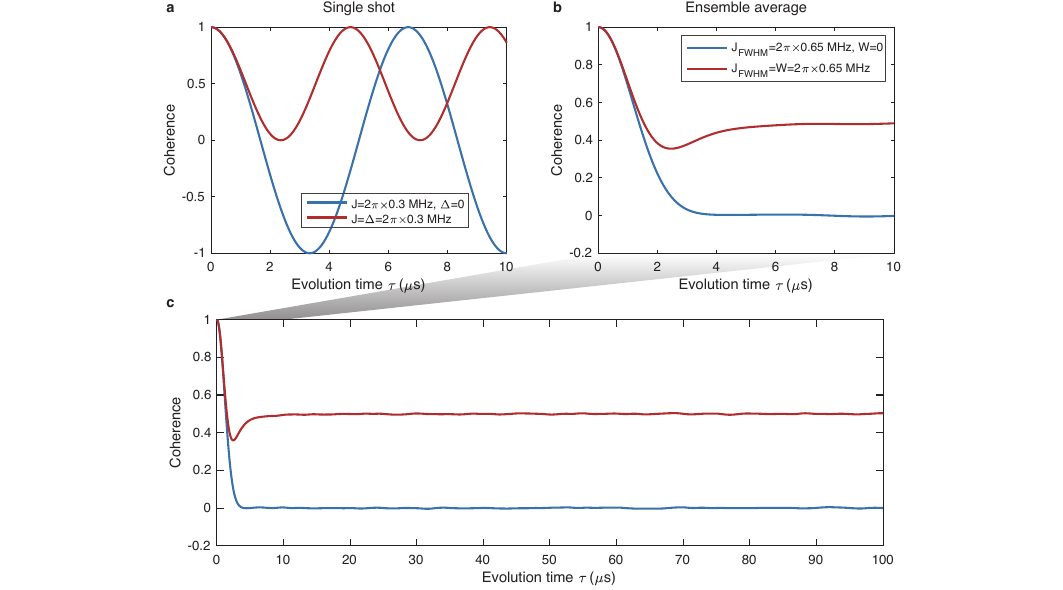}
    \caption{\textbf{Two-spin model with interaction strength $J$ and relative detuning $\Delta$.} Two spins are initially polarized along the $y$-axis, and coherence is defined as the remnant polarization along the $y$-axis after evolving for a duration $\tau$. \textbf{a,} Single realization (shot) dynamics of coherence for a fixed $J=2\pi\times 0.3$ MHz, but with $\Delta=0$ (blue) and $\Delta=J$ (red). \textbf{b,} Ensemble-averaged dynamics of coherence for different $J$ and $\Delta$ values. The distribution of $J$ is assumed to follow a Gaussian distribution with a full width at half maximum (FWHM) of $J_\text{FHWM} = 2\pi \times 0.65$ MHz. Two cases are considered for a distribution of relative detuning $\Delta$, which is also assumed to follow a Gaussian distribution with a FWHM of $W$: one with no disorder ($W=0$, blue) and another with disorder ($W=2\pi \times 0.65$ MHz, red). \textbf{c,} Longer-time evolution is simulated for \textbf{b}. The presence of on-site disorder gives rise to a non-zero offset in the late-time saturated coherence.}
    \label{twoatoms}
\end{figure}

For a pair of spins with interaction strength $J$ and relative detuning $\Delta$, we can derive the average spin polarization along the $\hat{y}$-axis at time $\tau$, $P(\tau)$, under the spin-echo sequence as follows:
\begin{equation} \label{Ptau}
P(\tau)=\frac{\Delta^2}{\Delta^2+J^2}+\frac{J^2}{\Delta^2+J^2}\cos(\frac{\sqrt{\Delta^2+J^2}}{2}\tau).
\end{equation} 
Recall that the polarization signal represents the spin coherence. For the resonant spins with $\Delta=0$, as in Model I, $P(\tau)=\cos(J\tau/2)$. This implies that the two spins oscillate between the $\hat{y}$ and $-\hat{y}$ directions with a frequency of $J/2$, similar to Rabi oscillation dynamics (blue curve, Fig.~\ref{twoatoms}a). Ensemble averaging over a distribution of $J$ leads to a decay in spin polarization, resulting from the incoherent averaging of sinusoidal oscillations with different frequencies (blue curve, Fig.~\ref{twoatoms}b).

However, when a non-zero detuning $\Delta \neq 0$ is introduced, the spin polarization signal develops a non-zero positive offset of $\frac{\Delta^2}{\Delta^2+J^2}$ with a reduced contrast of $\frac{J^2}{\Delta^2+J^2}$ (red curve, Fig.~\ref{twoatoms}a, and see Eq.~\ref{Ptau}). Similar to the earlier case, ensemble averaging over random distributions of $J$ and $\Delta$ leads to a decay in spin polarization (red curve, Fig.~\ref{twoatoms}b). Notably, we observe that the early-time decay rate at small $\tau$ remains the same between the no-disorder and disordered cases (Fig.~\ref{twoatoms}b). Specifically, the ensemble-averaged coherence decay rate at early times is given by
\begin{equation}
    \left \langle \frac{dP}{d\tau} \right \rangle = \left \langle -\frac{J^2}{2\sqrt{\Delta^2+J^2}} \sin(\frac{\sqrt{\Delta^2+J^2}}{2}\tau)  \right \rangle \approx -\frac{\langle J^2 \rangle}{4}\tau,
\end{equation}
which depends only on $\langle J^2 \rangle$, the expectation value of the interaction strength squared. Here, $\langle ... \rangle$ denotes ensemble averaging. 

At late times, however, the presence of on-site disorder gives rise to a non-zero offset in the late-time saturated coherence, which does not decay in this two-spin model (red curve, Fig.~\ref{twoatoms}c). This suggests that involving more ions is necessary to explain our experimental observation of a slow decay emerging in the late-time decoherence profile, as numerically demonstrated in Fig.~\ref{MB_simu_fig}b.

\subsection{Three-spin case} \label{secThreeSpin}
Here we introduce a third spin to analyze how it affects the coherence dynamics. First, we examine a case with no on-site disorder for simplicity, where the three-spin Hamiltonian is given by
\begin{equation} \label{threespinHam}
\hat{H} = J_0(\hat{S}_x^1 \hat{S}_x^2+ \hat{S}_y^1 \hat{S}_y^2) + J_1(\hat{S}_x^1 \hat{S}_x^3+ \hat{S}_y^1 \hat{S}_y^3)+J_2(\hat{S}_x^2 \hat{S}_x^3+ \hat{S}_y^2 \hat{S}_y^3).
\end{equation}
We assume that the third spin interacts only weakly with both spins at sites 1 and 2, namely, $|J_1|,|J_2| \ll |J_0|$. By employing second-order perturbation theory with the dominant Hamiltonian $\hat{H}_0 = J_0(\hat{S}_x^1 \hat{S}_x^2+ \hat{S}_y^1 \hat{S}_y^2)$ and the perturbing Hamiltonian $\hat{V} = J_1(\hat{S}_x^1 \hat{S}_x^3+ \hat{S}_y^1 \hat{S}_y^3)+J_2(\hat{S}_x^2 \hat{S}_x^3+ \hat{S}_y^2 \hat{S}_y^3)$, we can solve for the coherence dynamics of each spin under the spin-echo sequence. Without loss of generality, we focus on the coherence dynamics of spin 1, $P_1$, whose analytical expression is shown to be the sum of oscillations at different frequencies with a DC offset:
\begin{equation}
    P_1=P_1^\text{DC}+P_1^{\frac{J_1J_2}{J_0}}+P_1^{J_0 + \nu_0}+P_1^{\frac{J_0}{2}+\nu_1}+P_1^{\frac{J_0}{2}+\nu_2}+P_1^{\frac{J_0}{2} + \nu_3}++P_1^{\frac{J_0}{2} + \nu_4},
\end{equation}
where the individual terms on the right-hand side are defined as follows:
\begin{align}
    P_1^\text{DC} &= \frac{J_2}{2J_0}+\frac{3J_1^2+J_2^2+4J_1J_2}{2J_0^2} \label{DC} \\
    P_1^{\frac{J_1J_2}{J_0}} &=
    -\frac{1}{2}\left(\frac{J_2}{J_0}-\frac{J_2(J_1+J_2)}{J_0^2}\right)\cos\left(\frac{J_1J_2}{J_0}\tau\right) \label{J1J2} \\
    P_1^{J_0 + \nu_0} &= \frac{J_1(J_1-J_2)}{2J_0^2}\cos\left({(J_0+\nu_0)\tau}\right) \\
    P_1^{\frac{J_0}{2}+\nu_1} &= \frac{1}{2}\left(1-\frac{J_1+J_2}{2J_0}-\frac{13J_1^2+7J_2^2+12J_1J_2}{4J_0^2}\right)\cos\left({(\frac{J_0}{2}+\nu_1)\tau}\right) \\
    P_1^{\frac{J_0}{2}+\nu_2} &= \frac{1}{2}\left(1+\frac{J_1+J_2}{2J_0}-\frac{J_1^2+3J_2^2+4J_1J_2}{4J_0^2}\right)\cos\left({(\frac{J_0}{2}+\nu_2)\tau}\right) \\
    P_1^{\frac{J_0}{2} + \nu_3} &= \frac{1}{2}\left(\frac{J_1-J_2}{2J_0}+\frac{3(J_2^2-J_1^2)}{4J_0^2}\right)\cos\left({(\frac{J_0}{2}+\nu_3)\tau}\right) \\
    P_1^{J_0/2 + \nu_4}&=\frac{1}{2}
    \left(\frac{J_2-J_1}{2J_0}+\frac{J_1^2-J_2^2}{4J_0^2}\right)\cos\left({(\frac{J_0}{2}+\nu_4)\tau}\right)
\end{align}
where $\nu_0 = \frac{J_1^2+J_2^2}{2J_0}$, $\nu_1 = \frac{J_1^2+J_2^2+6J_1J_2}{4J_0}$, $\nu_2 = \frac{(J_1+J_2)^2}{4J_0}$, $\nu_3 = \frac{J_1^2+J_2^2-6J_1J_2}{4J_0}$, and $\nu_4 = \frac{(J_1-J_2)^2}{4J_0}$.

\begin{figure}[hbt!]
    \centering
    \includegraphics[width=\linewidth]{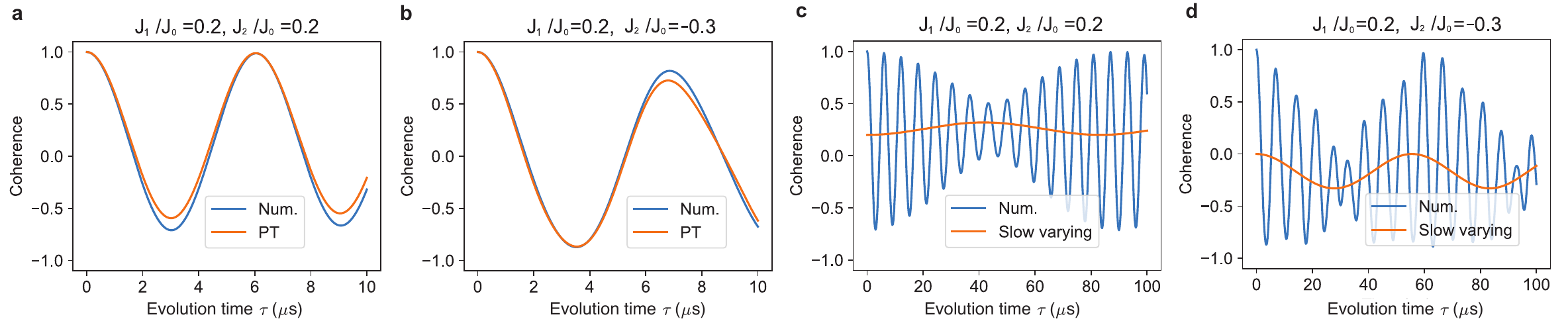}
    \caption{\textbf{Comparison between numerical simulation and perturbation theory for three resonant spins.} We monitor the short-time (\textbf{a, b}) and long-time (\textbf{c, d}) coherence dynamics of a single spin interacting strongly with its nearest-neighbor spin with strength $J_0$ and with a farther spin with strength $J_1$. The second and third spins interact with strength $J_2$. All spins are initially polarized along the $y$-axis and subsequently subject to resonant spin-exchange Hamiltonian dynamics~(Eq.~\ref{threespinHam}). In \textbf{a,c}, $J_1/J_0 = J_2/J_0 = 0.2$, while in \textbf{b,d}, $J_1/J_0 = 0.2, J_2/J_0 = -0.3$. In all cases, $J_0 = 2\pi\times0.3$ MHz is fixed. For the short-time dynamics (\textbf{a, b}), we compare the full numerical simulations (blue) with predictions by perturbation theory (orange). For the long-time dynamics (\textbf{c, d}), we show the full numerical simulations (blue) alongside the offset oscillation of the slow-frequency component predicted by perturbation theory (orange, Eqs.~\ref{DC} and \ref{J1J2}).}
    \label{threeatomcombine}
\end{figure}

As shown in Fig.~\ref{threeatomcombine}, the analytical perturbative solutions are validated by comparing them with numerical simulations, showing reasonable agreement. Interestingly, similar to the {\it disordered} two-spin case, the {\it resonant} three-spin scenario also exhibits an offset (Eq.~\ref{DC}) in coherence oscillation with reduced contrast, as the additional third spin prevents the spin-exchange dynamics between spins 1 and 2 from perfectly `rephasing' back to their original condition (Fig.~\ref{threeatomcombine}a,b). Moreover, we identify a very slow oscillation component with frequency $\frac{J_1 J_2}{J_0}$ (Eq.~\ref{J1J2}) in the three-spin model (Fig.~\ref{threeatomcombine}c,d), which results in the slow late-time relaxation upon ensemble averaging. The offset (Eq.~\ref{DC}), as well as the slow frequency component  (Eq.~\ref{J1J2}) appearing in the large-spin model, are attributed to the long tail observed in the late-time decoherence profile in Model II (see Fig.~3 of the main text).  

\begin{figure}[hbt!]
    \centering
    \includegraphics[width=0.6\linewidth]{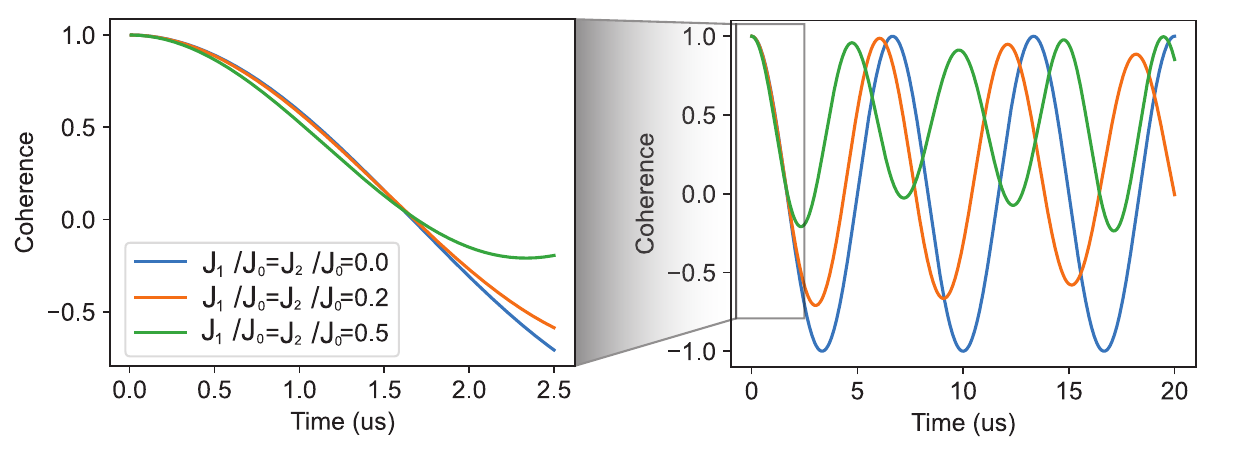}
    \caption{\textbf{Short-time behavior of the three-spin model for different interaction strengths.} Three different conditions of ($J_1/J_0$, $J_2/J_0$) are considered for numerical simulation, as indicated in the figure legend of the left panel. The blue trace with $J_1/J_0=J_2/J_0=0$ represents the resonant two-spin case. In all cases, $J_0 = 2\pi\times0.3$ MHz is fixed. The right panel shows that the three-spin cases (orange and green) exhibit reduced contrast with beating frequencies, due to the hindered perfect rephasing of coherence dynamics caused by the additional spin. Refer to Sec.~\ref{secThreeSpin} for discussion.}
    \label{diffJ0J1J2}
\end{figure}

Finally, the early-time coherence decay rate of spin 1 is primarily dominated by $J_0$, the strongest interaction strength between itself and spin 2, with a small correction due to its weak coupling to the third spin, $J_1$:
\begin{equation}
    \frac{dP_1}{d\tau} \approx -\frac{J_0^2+J_1^2}{4}\tau \approx -\frac{J_0^2}{4}\tau,
\end{equation}
when $\tau \ll 1/|J_0|$ and $J_0^2 \gg J_1^2$ (Fig.~\ref{diffJ0J1J2}).

\section{Additional experimental data}
In this section, we provide additional experimental data and corresponding analyses related to the experiments shown in Figures 4 and 5 of the main text.

\subsection{WAHUHA-echo measurement} \label{SecWAHUHA}
\begin{figure}[hbt!]
    \centering
    \includegraphics[width=\linewidth]{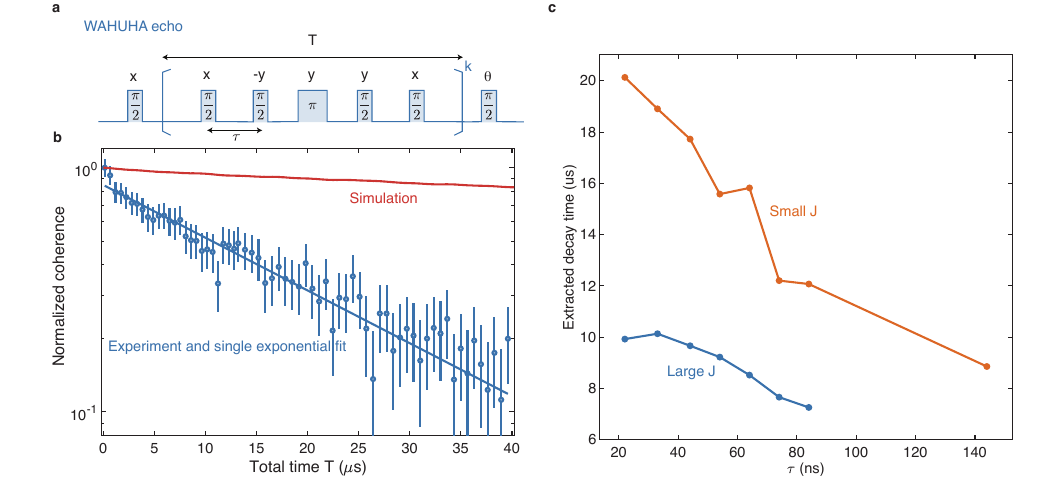}
    \caption{\textbf{WAHUHA-echo sequence measurement.} \textbf{a,} WAHUHA-echo sequence. The base pulse sequence consists of $\pi/2$ and $\pi$ pulses with a periodicity of $6\tau$, where $\tau$ denotes the time separation between the centers of two adjacent $\pi/2$ pulses. We maintain a spacing of $\tau - t_p$ between each pulse, where $t_p$ is the duration of the $\pi/2$ pulse. Note that the duration of a $\pi$ pulse is $2t_p$. The base sequence is repeated $k$ times to evolve over a total interrogation time $T$. The final $\pi/2$ analyzer pulse with a variable phase angle $\theta$ is employed for coherence extraction. \textbf{b,} Comparison of the WAHUHA-echo coherence dynamics between numerical simulation (red) and experiment (blue). The experiment is carried out under the small $J$ condition, with error bars representing the standard deviation of the data (see Sec.~\ref{cohe} for details). The simulation is conducted using the experimentally calibrated interaction and disorder strengths. The blue line is a single exponential fit to extract the $1/e$ decay time. \textbf{c,} The $1/e$ decay times of the WAHUHA-echo coherence measurements are shown as a function of the pulse separation $\tau$ for both the large $J$ (blue) and small $J$ (orange) cases.}
    \label{WAHUHA}
\end{figure}

In the WAHUHA-echo sequence~\cite{choi2020robust}, the base pulse sequence consists of $\pi/2$ and $\pi$ pulses with judiciously chosen rotation axes, which are repeatedly applied to the spin system. The base sequence has a periodicity of $6\tau$, where $\tau$ here represents the time separation between the centers of two adjacent $\pi/2$ pulses (Fig.~\ref{WAHUHA}a). To extend the interrogation up to the total time $T$, the base sequence with a fixed $\tau$ is repeated $k$ times, and coherence is measured stroboscopically at discrete times at the end of each repetition. Specifically, we sweep the rotation axis $\theta$ of the $\pi/2$ analyzer pulse at the end of the entire sequence, fit the resulting photoluminescence signal as a function of $\theta$, and extract the contrast (see Fig.~\ref{echodetails}b for details). To quantify the decoherence rates under the WAHUHA-echo sequence, we fit the coherence decay profile as a function of $T$ with a single exponential decay (blue line, Fig.~\ref{WAHUHA}b) and extract the $1/e$ decay times (Fig.~\ref{WAHUHA}c). The $1/e$ decay times under the WAHUHA-echo sequence are characterized as a function of the $\pi/2$ pulse separation $\tau$ for both the small $J$ and large $J$ cases (orange/blue markers, Fig.~\ref{WAHUHA}c).

From the theoretical perspective of Floquet Hamiltonian engineering, a shorter $\tau$ more effectively suppresses the contribution from higher $m^\text{th}$-order Hamiltonians, thereby leading to the effective Hamiltonian being dominated by the zeroth-order Hamiltonian, $\hat{H}_\text{eff} \approx \hat{H}_0$, in the resultant spin dynamics (Eq.~\ref{Heff}). Since the zeroth-order Hamiltonian of the WAHUHA-echo sequence is the isotropic Heisenberg interaction Hamiltonian, i.e., $\hat{H}_0 = \frac{2}{3}\hat{H}_\text{Heis}$, this enables us to protect the coherence of the polarized spin ensemble by effectively decoupling spin-spin interactions. 

We experimentally confirm that coherence is indeed better preserved as $\tau$ decreases, indicating the dominance of the zeroth-order average Hamiltonian in the Floquet-engineered many-body dynamics (Fig.~\ref{WAHUHA}c). However, in experiments, control imperfections due to finite pulse duration and rotation angle errors accumulate more for smaller $\tau$ because a larger number of pulses are needed to achieve a longer interrogation time $T$. We speculate that these imperfections led to the discrepancy between the numerical simulation (red line, Fig.~\ref{WAHUHA}b) and the experiment (blue markers, Fig.~\ref{WAHUHA}b), as well as the saturation of coherence improvement as $\tau$ approaches the finite pulse duration, i.e., $\tau \approx 20$~ns, for the large $J$ case (Fig.~\ref{WAHUHA}c). In our experiments, the maximum coherence times are attained when $\tau = 22$ ns and $\tau = 33$ ns for the small $J$ and large $J$ cases, respectively. We anticipate that coherence times could improve with the implementation of shorter pulse durations and more robust pulse sequences~\cite{choi2020robust}.

\subsection{Spin-locking measurement} \label{SecSL}
As presented in Figure 4b of the main text, the spin-locking sequence is employed to investigate how long the polarized spin ensemble remains stably locked along the $y$-axis under a continuous drive with strength $\Omega_y$. Specifically, under the spin-lock drive, the zeroth-order average Hamiltonian can be shown to be 
\begin{equation}
    \hat{H}_0 = \Omega_y \sum_i^N \hat{S}_y^i + \frac{1}{2} \sum_{ij, \; i>j}^N J_{ij}(\hat{S}_y^i \hat{S}_y^j+\vec{S^i} \cdot\vec{S^j}) = \hat{H}_\text{on-site}^y + \frac{1}{2} \hat{H}^y_\text{Ising} + \frac{1}{2} \hat{H}_\text{Heis}.
\end{equation}
Note that the zeroth-order Hamiltonian can be approximated as $\hat{H}_0 \approx \hat{H}_\text{on-site}^y = \Omega_y \sum_i^N \hat{S}_y^i$ when $\Omega_y \gg J_{ij}$. While in the leading order, the polarized spin ensemble can be protected against interaction-induced dephasing since the initial state is an eigenstate of $\hat{H}_0$, higher-order average Hamiltonians, as well as environmental noise from the spin bath creating AC noise at frequency $\Omega_y$, could induce dephasing of the spin-locked signal.

\begin{figure}[hbt!]
    \centering
    \includegraphics[width=\linewidth]{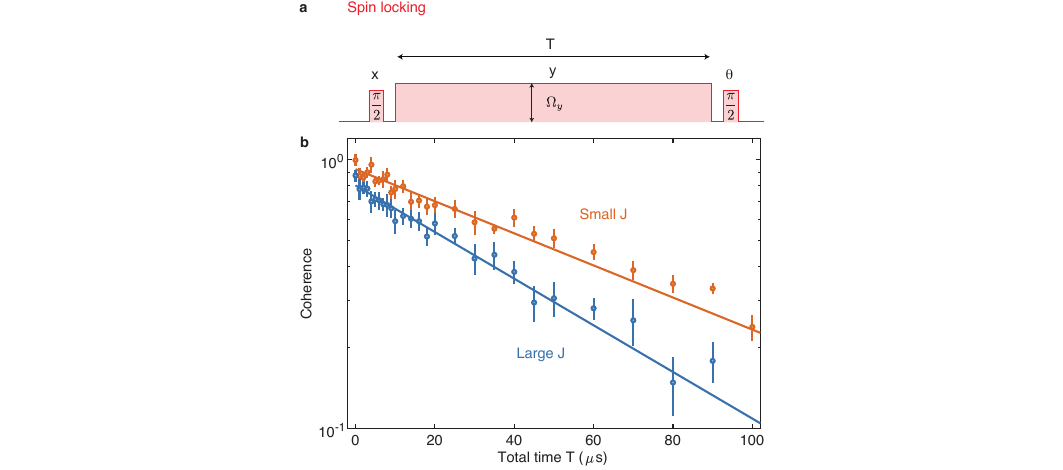}
    \caption{\textbf{Spin-locking sequence measurement.} \textbf{a,} Spin-locking sequence. After initializing the spin ensemble along the $y$-axis via the initial $\pi/2$ pulse, a continuous drive along the $y$-axis with strength $\Omega_y$ is applied for duration $T$ to lock the spin orientation. The final $\pi/2$ analyzer pulse with a variable phase angle $\theta$ is employed for coherence extraction. \textbf{b,} Decoherence dynamics of the spin ensemble under the spin-locking sequence for large $J$ (blue) and small $J$ (orange) cases. Error bars represent the standard deviation of the experimental data (see Sec.~\ref{cohe} for details). The solid lines are single exponential fits used to extract the $1/e$ decay times, $T_{1/e}$, yielding $T_{1/e} \approx 50~\mu$s and $T_{1/e} \approx 73~\mu$s for the large $J$ and small $J$ cases, respectively.}
    \label{SL_times}
\end{figure}

In Figure~\ref{SL_times}, we experimentally probe the decoherence dynamics of the spin-locking sequence for both small $J$ and large $J$ cases. The spin-lock Rabi frequency is fixed at $\Omega_y \approx 2\pi \times 10$ MHz, significantly greater than other rates in the system, including large $J = 2\pi \times 0.35$ MHz, small $J = 2\pi \times 0.19$ MHz, and $W = 2\pi \times 0.65$ MHz. The decay time traces are fitted with a single exponential function, yielding $1/e$ decay times of $\approx$$50~\mu$s and $\approx$$73~\mu$s for the large $J$ and small $J$ cases, respectively. We attribute the different decay times observed at different $J$ values to contributions from the aforementioned higher-order average Hamiltonians and varying levels of device heating induced by optical power during spin density tuning.

\subsection{Robustness of DTC phases to initial states}
\begin{figure}[hbt!]
    \centering
    \includegraphics[width=\linewidth]{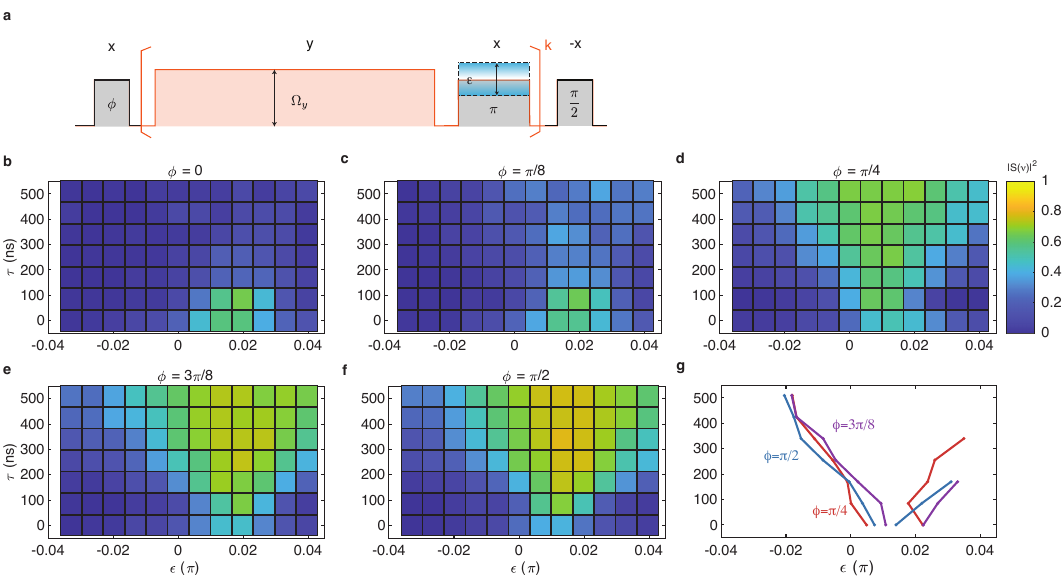}
    \caption{\textbf{Robustness of DTC phases to initial spin states} \textbf{a,} Floquet control pulse sequence for probing DTC phases with different initial spin orientations, controlled by the first pulse with a variable rotation angle $\phi$; otherwise, the sequence is the same as in Fig.~5a of the main text. \textbf{b-f,} DTC phase diagrams for different initial rotation angles ranging from $\phi=0$ to $\phi=\pi/2$. See the main text for details on the phase diagram reconstruction. \textbf{g,} Phase boundaries for \textbf{d-f} where the initial spin orientations are near the $y$-axis. The phase boundaries for each interaction time $\tau$ are determined by identifying a threshold in the value of $\epsilon$ where the subharmonic peak intensity at frequency $\nu=1/2$, $|S(\nu=1/2)|^2$, falls below 0.4, i.e., $|S(\nu=1/2)|^2 < 0.4$. We note a lateral 1\% offset in the swept angle $\epsilon$ at the center position of the DTC phase diagram, i.e., $\epsilon/\pi \approx 0.01$, attributed to an experimental calibration error of the rotation pulse.}
    \label{robustnessDTC}
\end{figure}

In Figure~\ref{robustnessDTC}, we investigate the robustness of discrete time-crystalline (DTC) phases relative to initial spin states. The Floquet pulse sequence used to investigate the robustness against initial states is similar to Fig.~5a of the main text. Here, however, we apply a variable angle $\phi$ ranging from $\phi=0$ to $\phi=\pi/2$ to the initialization pulse, preparing initial spin orientations along the $z$-axis and $y$-axis, respectively (Fig.~\ref{robustnessDTC}a). For each initial state specified by a rotation angle $\phi$, we sweep the perturbation strength,  $\epsilon$, and the interaction time, $\tau$, of the base control sequence to generate a DTC phase diagram (Fig.~\ref{robustnessDTC}b-f). Details on the construction procedures for DTC phase diagrams can be found in the main text.

We experimentally observe that when the initial spin orientation is near the $y$-axis, for instance, when $|\phi - \pi/2| \lesssim \pi/4$ (Fig.~\ref{robustnessDTC}d-f), the DTC phases robustly exhibit characteristic linear phase boundaries (Fig.~\ref{robustnessDTC}g). We observe that the slope of the phase boundaries remains consistent for different $\phi$ values, implying the robustness of the observed subharmonic DTC signals relative to the initial spin states. 

In contrast, when the initial spin orientation is near the $z$-axis (Fig.~\ref{robustnessDTC}a,b), the subharmonic DTC signals are visible only within a small region near $\tau \approx 0$ and $\epsilon \approx 0$, indicating rapid decoherence under the applied Floquet sequence. We attribute this to the degradation of the underlying spin-locking performance, as the spin ensemble can be dephased by the dominant on-site field along the $y$-axis when the spin orientation deviates too much from the spin-locking field direction. In other words, when the initial spins are aligned close to the $z$-axis, the continuous drive induces Rabi precession instead of spin locking.

\clearpage
\let\oldthebibliography=\thebibliography
\let\oldendthebibliography=\endthebibliography
\renewenvironment{thebibliography}[1]{
    \oldthebibliography{#1}
    \setcounter{enumiv}{50}
}{\oldendthebibliography}
\bibliography{mainSI}
\bibliographystyle{naturemag}